# On some counting problems for semi-linear sets *

Flavio D'Alessandro
Dipartimento di Matematica,
Università di Roma "La Sapienza"
Piazzale Aldo Moro 2, 00185 Roma, Italy.
e-mail: dalessan@mat.uniroma1.it,
http://www.mat.uniroma1.it/people/dalessandro

Benedetto Intrigila
Dipartimento di Matematica,
Università di Roma "Tor Vergata",
via della Ricerca Scientifica, 00133 Roma, Italy.
e-mail intrigil@mat.uniroma2.it,

Stefano Varricchio
Dipartimento di Matematica,
Università di Roma "Tor Vergata",
via della Ricerca Scientifica, 00133 Roma, Italy.
e-mail varricch@mat.uniroma2.it,
http://mat.uniroma2.it/~varricch

Stefano Varricchio suddenly passed away on August 20th 2008, shortly after this paper has been completed. We will remember Stefano as the best friend of us and as an outstanding researcher. Working with Stefano was always an enthusiastic experience both for his beautiful and original ideas and for his scientific *rigueur*.

**Abstract**

Let $X$ be a subset of $\mathbb{N}^t$ or $\mathbb{Z}^t$. We can associate with $X$ a function $\mathcal{G}_X : \mathbb{N}^t \longrightarrow \mathbb{N}$ which returns, for every $(n_1, \ldots, n_t) \in \mathbb{N}^t$, the number $\mathcal{G}_X(n_1, \ldots, n_t)$ of all vectors $x \in X$ such that, for every $i = 1, \ldots, t$, $|x_i| \leq n_i$. This function is called the *growth function* of $X$. The main result of this paper is that the growth function of a semi-linear set of $\mathbb{N}^t$ or $\mathbb{Z}^t$ is a box spline. By using this result and some theorems on semi-linear sets, we give a new proof of combinatorial flavour of a well-known theorem by

*This work was partially supported by MIUR project "Aspetti matematici e applicazioni emergenti degli automi e dei linguaggi formali". The first author acknowlegdes the partial support of fundings "Facoltà di Scienze MM. FF. NN. 2007" of the University of Rome "La Sapienza".

Dahmen and Micchelli on the counting function of a system of Diophantine linear equations.

# 1 Introduction

We study some counting problems on semi-linear sets of the free commutative monoid $\mathbb{N}^t$ and of the free commutative group $\mathbb{Z}^t$. The notion of semi-linear set has been widely investigated in the past because it plays an important role in the study of several problems of Mathematics and Computer Science. In this context, it is worth to mention a well-known theorem by S. Ginsburg and E. H. Spanier [10] (see also [9, 13]) that establishes a strong connection between semi-linear sets, rational sets on $\mathbb{N}$ and on $\mathbb{Z}$ and Presburger definable sets. In [8], Eilenberg and Schützenberger extended the connection between semi-linear sets and rational sets to every finitely generated commutative monoid.

It is now convenient to introduce two concepts that are central in our work: those of *quasi-polynomial* and *box-spline*. A quasi-polynomial is a map $F : \mathbb{N}^t \to \mathbb{N}$, defined by a finite family of polynomials in $t$ variables $x_1, \ldots, x_t$, with rational coefficients:

$$\{p_{(d_1,d_2,\cdots,d_t)} \mid d_1, \ldots, d_t \in \mathbb{N},\ 0 \leq d_i < d\},$$

such that every polynomial of the family is indexed by a vector $(d_1, d_2, \cdots, d_t)$ whose components are remainders of a fixed positive integer $d$. Then, for every $(n_1, \ldots, n_t) \in \mathbb{N}^t$, the value of $F$ computed at $(n_1, \ldots, n_t)$ is given by:

$$F(n_1, \ldots, n_t) = p_{(d_1,d_2,\cdots,d_t)}(n_1, \ldots, n_t),$$

where, for every $i = 1, \ldots, t$, $d_i$ is the remainder of the division of $n_i$ by $d$.

In this paper, we shall use a particular kind of box-spline specified as a map $F : \mathbb{N}^t \to \mathbb{N}$ such that there exist a partition of $\mathbb{N}^t$ into a finite number of polyhedral conic regions $R_1, \ldots, R_s$, determined by hyperplanes, through the origin, with rational equations, and a finite number of quasi-polynomials $p_1, \ldots p_s$, where, for any $(n_1, \ldots, n_t)$ one has:

$$F(n_1, \ldots, n_t) = p_j(n_1, \ldots, n_t),$$

where $j$ is such that $(n_1, \ldots, n_t) \in R_j$.
In the sequel, we shall use the notion of box-spline in the sense specified above. One interesting application of these notions was first given by E. T. Bell in [1], by proving that the counting map of a diophantine linear equation is a quasi-polynomial. Later, this result was extended in a paper by Dahmen and Micchelli [5], where it has been proved that the counting function of a Diophantine system of linear equations can be described by a set of quasi-polynomials, under suitable conditions on the matrix of the system. Recently this result has been object of further investigations in [6, 7, 15] where important theorems on the algebraic and combinatorial structure of partition functions have been obtained. In formal language theory, the notion of quasi-polynomial has been used to describe the counting and the growth function of regular languages [14] and of bounded context-free languages [3, 4]. In particular, in [4], a result perhaps of some interest is that the Parikh counting function of a bounded context-free language is a box-spline. Moreover, starting from the grammar that generates such a language, one can effectively construct a box-spline that describes its Parikh counting function.

In this paper, we introduce and study the notion of *growth function* of semi-linear sets both over $\mathbb{N}$ as well as over $\mathbb{Z}$. More precisely, we define the *growth function* of a subset $X$ of $\mathbb{N}^t$ or $\mathbb{Z}^t$, $t \geq 0$ as the function $f_X : \mathbb{N}^t \longrightarrow \mathbb{N}$ which associates with non negative integers $n_1, \ldots, n_t$, the number $f_X(n_1, \ldots, n_t)$ of all the elements $(n_1, \ldots, n_t) \in X$ such that:

$$\begin{cases} |x_1| & \leq & n_1 \\ \cdot & & \cdot \\ \cdot & & \cdot \\ \cdot & & \cdot \\ |x_t| & \leq & n_t. \end{cases}$$

The function $f_X$ seems to be a natural way to count the number of points of $X$. Indeed, in case $X$ lies in $\mathbb{N}^t$ (resp. $\mathbb{Z}^t$), it counts the number of points of $X$ lying in larger and larger hyper-parallelepipeds, starting from the origin of $\mathbb{N}^t$ (resp. centered at the origin of $\mathbb{Z}^t$). In this paper, by using the combinatorial techniques developped in [4], we prove that the growth function of every semi-linear set of $\mathbb{N}^t$ or $\mathbb{Z}^t$, $t \geq 0$, is always a box spline. Furthermore, this box-spline can be effectively computed. Moreover, by using such techniques and some deep results from the theory of semi-linear sets, we give a new proof of the theorem of Dahmen and Micchelli mentioned above.

The paper is organized as follows. In Section 2, after introducing some basic notions and results, we recall the definition of growth function of a semi-linear set. In Section 3, we give a combinatorial proof of the fact that the counting function of a system of diophantine linear equations with coefficients in $\mathbb{N}$ is a box spline. This will be a crucial tool to prove the results of the subsequent sections. We remark that the proof of the result mentioned above has been already published in [4] and we reproduce it here for the sake of completeness. In Section 4, we study the growth functions of semi-linear sets and, as the main result of this section, we prove that such functions are box-spline. Finally, in Section 5, we give a proof of combinatorial flavour of the quoted theorem by Dahmen and Micchelli.

# 2 Preliminaries and basic definitions

The aim of this section is to recall some results about semi-linear sets of the free commutative monoid and the free commutative group. For this purpose, we follow [2]. The free abelian monoid and the free abelian group on $k$ generators are respectively identified with $\mathbb{N}^k$ and $\mathbb{Z}^k$ with the usual additive structure. The operation of addition is extended from elements to subsets: if $X, Y \subseteq \mathbb{N}^k$ (resp. $X, Y \subseteq \mathbb{Z}^k$), $X + Y \subseteq \mathbb{N}^k$ (resp. $X + Y \subseteq \mathbb{Z}^k$) is the set of all sum $x + y$, where $x \in X, y \in Y$. It might be convenient to consider the elements of $\mathbb{N}^k$ and $\mathbb{Z}^k$ as vectors of the $\mathbb{Q}^k$-vector space $\mathbb{Q}^k$. Given $v$ in $\mathbb{N}^k$ or in $\mathbb{Z}^k$, the expression $\mathbb{N}v$ stands for the subset of all elements $nv$, where $n \in \mathbb{N}$. This expression can be extended to $\mathbb{Z}v$, whenever $v$ is in $\mathbb{Z}^k$. Let $B = \{b_1, \ldots, b_n\}$ be a finite subset of $\mathbb{Z}^k$. Then we denote by $B^\oplus$ the submonoid of $\mathbb{N}^k$ generated by $B$, that is

$$B^\oplus \ = \ b_1^\oplus + \cdots + b_n^\oplus \ = \ \{m_1 b_1 + \cdots + m_n b_n \ \mid \ m_i \in \mathbb{N}\}.$$

In the sequel, the symbol $\mathbb{K}$ stands for $\mathbb{N}$ when it concerns the free abelian monoid $\mathbb{N}^k$ and for $\mathbb{Z}$ when it concerns the free abelian group $\mathbb{Z}^k$. The following definitions are useful.

**Definition 1** *Let $X$ be a subset of $\mathbb{Z}^k$ (resp. $\mathbb{N}^k$). Then*

1. *$X$ is* $\mathbb{K}$-linear *if it is of the form*

$$a + \sum_{i=1}^{n} \mathbb{K} b_i, \ \ a, b_i \in \mathbb{Z}^k, \ \ (\textit{resp.}\mathbb{N}^k), \ \ i = 1, \ldots, n;$$

2. *$X$ is* $\mathbb{K}$-simple *if the vectors $b_i$ are linearly independent in $\mathbb{Q}^k$,*

3. *$X$ is* $\mathbb{K}$-semi-linear *if $X$ is a finite union of $\mathbb{K}$-linear sets;*

4. *$X$ is* semi-simple *if $X$ is a finite disjoint union of $\mathbb{K}$-simple sets.*

**Remark 1** In the definition of simple set, the vector $a$ and those of the set $\{b_1, \ldots, b_n\}$ shall be called a *representation* of $X$.

There exists a classical and important connection between the concept of semi-linear set and the *Presburger arithmetic*. Denote by $\mathcal{Z} = \langle \mathbb{Z}; =; <; +; 0; 1 \rangle$ and by $\mathcal{N} = \langle \mathbb{N}; =; +; 0; 1 \rangle$ respectively the *standard* and the *positive Presburger arithmetic*. Given a subset $X$ of $\mathbb{N}^k$ (resp. $\mathbb{Z}^k$), we say that $X$ is *first-order definable* in $\mathcal{N}$ (resp. $\mathcal{Z}$), or a *Presburger set* of $\mathbb{N}^k$ (resp. $\mathbb{Z}^k$), if

$$X = \{(x_1, \ldots, x_k) \mid P(x_1, \ldots, x_k) \text{ is true}\},$$

where $P$ is a Presburger formula (with at most $k$ free variables) over $\mathbb{N}$ (resp. $\mathbb{Z}$). Ginsburg ans Spanier proved in [10] the following result for $\mathbb{N}^k$ that holds for $\mathbb{Z}^k$ also.

**Theorem 1** *(Ginsburg and Spanier, 1966) Given a subset $X$ of $\mathbb{N}^k$ (resp. $\mathbb{Z}^k$), the following assertions are equivalent:*

1. *$X$ is first-order definable in $\mathcal{N}$ (resp. $\mathcal{Z}$);*

2. *$X$ is $\mathbb{N}$-semilinear in $\mathbb{N}^k$ (resp. $\mathbb{Z}^k$);*

*3. X is* $\mathbb{N}$*-semisimple in* $\mathbb{N}^k$ *(resp.* $\mathbb{Z}^k$*).*

**Remark 2** A $\mathbb{Z}$-semilinear set of $\mathbb{Z}^k$ is always $\mathbb{N}$-semilinear in $\mathbb{Z}^k$.

**Remark 3** Theorem 1 is effective. Indeed, one can effectively represent a $\mathbb{N}$-semi-linear set $X$ as a semi-simple set. More precisely, one can effectively construct a finite family $\{V_i\}$ of finite sets of vectors such that the vectors in $V_i$ form a representation of a simple set $X_i$ and $X$ is the disjoint union of the sets $X_i$.

**Remark 4** Given a monoid $M$, a subset of $M$ is *rational* if it is obtained from finite subsets of $M$ by applying finitely many times the rational operations, that is, the set union, the product, and the *Kleene closure* operator. Obviously, a semi-linear set of $\mathbb{N}^k$ or $\mathbb{Z}^k$ is rational but one can prove that the opposite is true (see [13]). Therefore, Conditions 2 and 3 of Theorem 1 and the property of rationality are equivalent in $\mathbb{N}^k$ or $\mathbb{Z}^k$. We recall that the previous equivalence has been proven in the larger context of finitely generated commutative monoids by Eilenberg and Schützenberger [8].

The following definition is important.

**Definition 2** *Let $X$ be a subset of $\mathbb{Z}^t$. We associate with $X$ a function*

$$\mathcal{G}_X : \mathbb{N}^t \longrightarrow \mathbb{N}$$

*defined as: for every $(n_1, ..., n_t) \in \mathbb{N}^t$, $\mathcal{G}_X(n_1, ..., n_t)$ is the number of elements $(x_1, ..., x_t) \in X$ such that:*

$$\left\{ \begin{array}{ccc} |x_1| & \leq & n_1 \\ \cdot & & \cdot \\ \cdot & & \cdot \\ \cdot & & \cdot \\ |x_t| & \leq & n_t. \end{array} \right. \qquad (1)$$

*The function $\mathcal{G}_X$ is called the* growth function *of $X$.*

# 3 Counting the non-negative solutions of systems of Diophantine linear equations with positive coefficients.

As we have said in the introduction of this paper, our main goal is to give an exact description of growth functions of semi-linear sets of $\mathbb{Z}^k$. For this purpose, now we develop a combinatorial tool to deal with such functions. This tool concerns systems of Diophantine linear equations. We start by considering systems of Diophantine linear equations with positive coefficients. Given a system of this kind, we can define a function that returns, for any vector of non homogenuous terms, the number of non-negative solutions of the system. We will prove that this function is a box-spline.

**Lemma 1** *Let $q(x_1, \ldots, x_t, x)$ be a polynomial in $t+1$ variables with rational coefficients and let*

$$F : \mathbb{N}^t \times \{\{-1\} \cup \mathbb{N}\} \longrightarrow \mathbb{Q}$$

*be the map defined as:*

$$F(x_1, \ldots, x_t, x) = \begin{cases} \sum_{\lambda=0,\ldots,x} q(x_1, \ldots, x_t, \lambda) & x \geq 0, \\ 0 & x = -1. \end{cases}$$

*There exists a polynomial $p(x_1, \ldots, x_t, x)$ in $t+1$ variables with rational coefficients such that, for every $(x_1, \ldots, x_t, x) \in \mathbb{N}^t \times \{\{-1\} \cup \mathbb{N}\}$, one has:*

$$F(x_1, \ldots, x_t, x) = p(x_1, \ldots, x_t, x).$$

*Proof.* Write $q(x_1, \ldots, x_t, x)$ as:

$$a_0 + a_1 x + \cdots + a_n x^n, \tag{2}$$

where, for every $i = 0, \ldots, n$, $a_i$ is a suitable polynomial in the variables $x_1, \ldots, x_t$ with rational coefficients. By Eq. (2), if $x \geq 0$, for every $x_1, \ldots, x_t \in \mathbb{N}$, one has:

$$F(x_1, \ldots, x_t, x) = \sum_{\lambda=0,\ldots,x} q(x_1, \ldots, x_t, \lambda) = \sum_{j=0,\ldots,n} \left( a_j \cdot \sum_{\lambda=0,\ldots,x} \lambda^j \right). \tag{3}$$

On the other hand, by using a standard argument (*cf* [4], Appendix), one can prove that, for any $j \in \mathbb{N}$, there exists a polynomial $p_j(x)$ with rational coefficients in one variable $x$ such that:

3.1) for any $x \in \mathbb{N}$, $p_j(x) = \sum_{\lambda=0,\ldots,x} \lambda^j$.

3.2) $p_j(-1) = 0$.

For any $j = 0, \ldots, n$, let $p_j$ be the polynomial defined above and let $p = p(x_1, \ldots, x_t, x)$ be the polynomial defined as:

$$p = \sum_{j=0,\ldots,n} a_j p_j.$$

Then by Eq. (3.2), one has $p(x_1, \ldots, x_t, -1) = 0$. Moreover, for every $x \geq 0$, by Eq. (3) and (3.1), one has:

$$F(x_1, \ldots, x_t, x) = \sum_{j=0,\ldots,n} \left( a_j \cdot \sum_{\lambda=0,\ldots,x} \lambda^j \right) = \sum_{j=0,\ldots,n} a_j p_j(x) = p(x_1, \ldots, x_t, x).$$

The proof is thus complete. □

**Definition 3** *A map $F : \mathbb{N}^t \longrightarrow \mathbb{N}$ is said to be a* quasi-polynomial *if there exist $d \in \mathbb{N}$, $d \geq 1$, and a family of polynomials in $t$ variables with rational coefficients:*

$$\{p_{(d_1,d_2,\cdots,d_t)} \ \mid \ d_1, \ldots, d_t \in \mathbb{N},\ 0 \leq d_i < d\},$$

*where, for every $(x_1, \ldots, x_t) \in \mathbb{N}^t$, if $d_i$ is the remainder of the division of $x_i$ by $d$, one has:*

$$F(x_1, \ldots, x_t) = p_{(d_1,d_2,\cdots,d_t)}(x_1, \ldots, x_t).$$

*The number $d$ is called the* period *of $F$.*

To simplify the notation, the polynomial $p_{(d_1,d_2,\cdots,d_t)}$ is denoted $p_{d_1 d_2 \cdots d_t}$.

**Definition 4** *Let $F : \mathbb{N}^t \longrightarrow \mathbb{N}$ be a map. Given a subset $C$ of $\mathbb{N}^t$, $F$ is said to be a* quasi-polynomial over $C$ *if there exists a quasi-polynomial $q$, such that $F(x) = q(x)$, for any $x \in C$.*

**Lemma 2** *The sum of a finite family of quasi-polynomials is a quasi-polynomial.*

*Proof.* It suffices to prove the claim for two quasi-polynomials. Let $f_1, f_2 : \mathbb{N}^t \longrightarrow \mathbb{N}$ be quasi-polynomials of periods $d_1, d_2$ respectively and let

$$\{p_{a_1 \cdots a_t} \ \mid \ \forall\, i = 0, \ldots, t,\ 0 \leq a_i \leq d_1 - 1\}, \ \text{ and}$$

$$\{q_{b_1 \cdots b_t} \ \mid \ \forall\, i = 0, \ldots, t,\ 0 \leq b_i \leq d_2 - 1\}$$

be the families of polynomials that define $f_1$ and $f_2$ respectively. Define a new quasi-polynomial $f$ as follows. Take $d = d_1 d_2$ as the period of $f$ and, for every $(c_1, \ldots, c_t) \in \{0, 1, \ldots, d-1\}^t$, take

$$f_{c_1 \cdots c_t} = p_{a_1 \cdots a_t} \ + \ q_{b_1 \cdots b_t},$$

where, for any $i = 1, \ldots, t$, $a_i$ and $b_i$ are the remainders of the division of $c_i$ by $d_1$ and $d_2$ respectively. It is easily checked that the quasi-polynomial $f$ is the sum of $f_1$ and $f_2$. Indeed, if $x = (x_1, \ldots, x_t) \in \mathbb{N}^t$ and, for every $i = 1, \ldots, t$, $x_i \ \equiv \ c_i \mod d$, then one has

$$c_i \ \equiv \ a_i \mod d_1 \ \iff \ x_i \ \equiv \ a_i \mod d_1$$

$$c_i \ \equiv \ b_i \mod d_2 \ \iff \ x_i \ \equiv \ b_i \mod d_2.$$

Therefore, if $x = (x_1, \ldots, x_t) \in \mathbb{N}^t$ and $x_i \equiv c_i \bmod d$, then we have:

$$f(x) = \ f_{c_1 \cdots c_t}(x) \ = \ p_{a_1 \cdots a_t}(x) \ + \ q_{b_1 \cdots b_t}(x) \ = \ f_1(x) \ + \ f_2(x).$$

The claim is thus proved. □

**Lemma 3** *Let $F : \mathbb{N}^t \longrightarrow \mathbb{N}$ be a map, $d$ be a positive integer, and $C$ be a subset of $\mathbb{N}^t$. If there exists a family of quasi-polynomials $\{F_{d_1 d_2 \cdots d_t} \ \mid \ d_1, \ldots, d_t \in \mathbb{N},\ 0 \leq d_i < d\}$, such that , for every $(x_1, \ldots, x_t) \in C$, with $x_i \equiv d_i \bmod d$, one has: $F(x_1, \ldots, x_t) = F_{d_1 d_2 \cdots d_t}(x_1, \ldots, x_t)$, then $F$ is a quasi-polynomial over $C$.*

*Proof.* Let $k$ be the least common multiple of $d$ and of the periods of the quasi-polynomials of the set $\{F_{d_1 d_2 \cdots d_t} \mid d_1, \ldots, d_t \in \mathbb{N},\ 0 \leq d_i < d\}$. Let $(r_1, \ldots, r_t)$ be a tuple of $\{0, 1, \ldots k-1\}^t$ and let $(x_1, \ldots, x_t) \in C$ be such that, for every $i = 1, ..., t$, $x_i \equiv r_i \bmod k$. Then one can check that $F(x_1, \ldots, x_t) = q(x_1, \ldots, x_t)$ where $q$ is a polynomial uniquely determined by $(r_1, \ldots, r_t)$. Indeed, one can first observe that the tuple $(r_1, \ldots, r_t)$ uniquely determines, for every $i = 1, ..., t$, the remainder $d_i$ of the division of $x_i$ by $d$ since $d_i \equiv r_i \bmod d$. By hypothesis, one has $F(x_1, \ldots, x_t) = F_{d_1 d_2 \cdots d_t}(x_1, \ldots, x_t)$. Since $k$ is a multiple of the period of the quasi-polynomial $F_{d_1 d_2 \cdots d_t}$, the tuple $(r_1, \ldots, r_t)$ also determines a polynomial $q$ in the family of polynomials associated with $F_{d_1 d_2 \cdots d_t}$, such that $F(x_1, \ldots, x_t) = F_{d_1 d_2 \cdots d_t}(x_1, \ldots, x_t) = q(x_1, ..., x_t)$. The proof is thus complete. □

**Lemma 4** *Let $\lambda : \mathbb{N}^t \longrightarrow \mathbb{Q}$ be a map such that, for any $(x_1, \ldots, x_t) \in \mathbb{N}^t$,*

$$\lambda(x_1, \ldots, x_t) = b_1 x_1 + \cdots + b_t x_t,$$

*where $b_1, \ldots, b_t$ are given rational coefficients. Let $C, C'$ be subsets of $\mathbb{N}^t$ and let $a_1, \ldots, a_t$ be non negative integers such that the following properties are satisfied: for any $(x_1, \ldots, x_t) \in C$, one has*

- $\lambda(x_1, \ldots, x_t) \geq 0$,
- *if $\lambda \in \mathbb{N}$ and $\lambda < \lambda(x_1, \ldots, x_t)$, then $(x_1 - \lambda a_1, x_2 - \lambda a_2, \ldots, x_t - \lambda a_t) \in C'$.*

*Let $p$ be a quasi-polynomial over $C'$ and define the map $F$ as:*

$$F(x_1, \ldots, x_t) = \sum_{0 \leq \lambda < \lambda(x_1, \ldots, x_t)} p(x_1 - \lambda a_1, x_2 - \lambda a_2, \ldots, x_t - \lambda a_t).$$

*Then $F$ is a quasi-polynomial over $C$.*

*Proof.* Let $d \geq 1$ be the period of the quasi-polynomial $p$. Let $p_{d_1 d_2 \cdots d_t}$, with $0 \leq d_i \leq d-1$, be the polynomials defining $p$. Consider the set of integers $\mu$:

$$0 \leq \mu \leq \lceil \lambda(x_1, \ldots, x_t) \rceil - 1,$$

and consider on it the partition:

$$F_0(x_1, \ldots, x_t) \cup F_1(x_1, \ldots, x_t) \cup \cdots \cup F_{d-1}(x_1, \ldots, x_t), \tag{4}$$

defined as: for any $j = 0, \ldots, d-1$:

$$\mu \in F_j(x_1, \ldots, x_t) \iff 0 \leq \mu \leq \lceil \lambda(x_1, \ldots, x_t) \rceil - 1 \text{ and } \mu \equiv j \pmod{d}.$$

By Eq. (4), for every $(x_1, \ldots, x_t) \in C$, we have:

$$F(x_1, \ldots, x_t) = \sum_{j=0}^{d-1} S_j(x_1, \ldots, x_t), \tag{5}$$

where, for any $j = 0, \dots, d-1$:

$$S_j(x_1, \dots, x_t) = \sum_{\mu \in F_j(x_1, \dots, x_t)} p(x_1 - \mu a_1, x_2 - \mu a_2, \dots, x_t - \mu a_t).$$

Now we prove that, for any $j = 0, \dots, d-1$, $S_j$ is a quasi-polynomial over $C$.

Let us fix a tuple $(d_1, \dots, d_t) \in \{0, 1, \dots d-1\}^t$. Let $(x_1, \dots, x_t)$ be such that $x_i \equiv d_i \bmod d$. Then for any $\mu \in F_j(x_1, \dots, x_t)$ one has $x_i - \mu a_i \equiv d_i - j a_i \bmod d$. Now set $q = p_{c_1 c_2 \cdots c_t}$, where $(c_1, \dots, c_t) \in \{0, 1, \dots d-1\}^t$ and $c_i \equiv d_i - j a_i \bmod d$. One has

$$S_j(x_1, \dots, x_t) = \sum_{\mu \in F_j(x_1, \dots, x_t)} q(x_1 - \mu a_1, x_2 - \mu a_2, \dots, x_t - \mu a_t).$$

On the other side, by Eq. (4), one easily checks:

$$F_j(x_1, \dots, x_t) = \left\{ j + d\mu \in \mathbb{N} \;\middle|\; 0 \leq \mu \leq \left\lfloor \frac{\lceil \lambda(x_1, \dots, x_t) \rceil - 1 - j}{d} \right\rfloor \right\}.$$

It is important to remark that, in the formula above, if $\lambda(x_1, \dots, x_t) = 0$, then

$$\frac{\lceil \lambda(x_1, \dots, x_t) \rceil - 1 - j}{d} < 0.$$

In this case, since for any $0 \leq j \leq d-1$, $|-1-j| \leq d$, one has

$$\left\lfloor \frac{-1-j}{d} \right\rfloor = -1.$$

Hence $\lambda(x_1, \dots, x_t) = 0$ implies that $F_j(x_1, \dots, x_t)$ is the empty set and thus $S_j$ is the null map. Let us consider the map

$$S : \mathbb{N}^t \times \{\{-1\} \cup \mathbb{N}\} \longrightarrow \mathbb{N},$$

where $S(x_1, \dots, x_t, x)$ is defined as:

$$\begin{cases} \sum_{\mu=0}^{x} q(x_1 - (j + \mu d)a_1, x_2 - (j + \mu d)a_2, \dots, x_t - (j + \mu d)a_t) & x \geq 0 \\ 0 & x = -1. \end{cases}$$

By the definition of the map $S$, for every $(x_1, \dots, x_t) \in C$, with $x_i \equiv d_i \bmod d$, one has:

$$S_j(x_1, \dots, x_t) = S\left( x_1, \dots, x_t, \left\lfloor \frac{\lceil \lambda(x_1, \dots, x_t) \rceil - 1 - j}{d} \right\rfloor \right). \qquad (6)$$

Therefore, by applying Lemma 1 to $S$, there exists a polynomial $Q(x_1, \dots, x_t, x)$ such that, for any $(x_1, \dots, x_t, x) \in \mathbb{N}^t \times \{\{-1\} \cup \mathbb{N}\}$,

$$S(x_1, \dots, x_t, x) = Q(x_1, \dots, x_t, x), \qquad (7)$$

hence, from Eqs. (6) and (7), for every $(x_1, \ldots, x_t) \in C$, with $x_i \equiv d_i \bmod d$, one has

$$S_j(x_1, \ldots, x_t) = Q(x_1, \ldots, x_t, \left\lfloor \frac{\lceil \lambda(x_1, \ldots, x_t) \rceil - 1 - j}{d} \right\rfloor). \tag{8}$$

By (*cf* [4], Appendix), one has that:

$$\left\lfloor \frac{\lceil \lambda(x_1, \ldots, x_t) \rceil - 1 - j}{d} \right\rfloor \tag{9}$$

is a quasi-polynomial in the variables $x_1, \ldots, x_t$. Therefore, by Eq. (8) , $S_j$ coincides with a quasi-polynomial on the set of points $(x_1, \ldots, x_t) \in C$, with $x_i \equiv d_i \bmod d$. Obviously this fact holds for any $(d_1, \ldots, d_t) \in \{0, 1, \ldots d-1\}^t$ and, by Lemma 3, $S_j$ is a quasi-polynomial over $C$. Finally, the fact that $F$ is a quasi-polynomial over $C$ follows from Eq. (5) by using Lemma 2. □

**Lemma 5** *Let $\lambda : \mathbb{N}^t \longrightarrow \mathbb{Q}$ be a map such that, for any $(x_1, \ldots, x_t) \in \mathbb{N}^t$,*

$$\lambda(x_1, \ldots, x_t) = b_1 x_1 + \cdots + b_t x_t,$$

*where $b_1, \ldots, b_t$ are given rational coefficients.*

*Let $C, C'$ be subsets of $\mathbb{N}^t$ and let $a_1, \ldots, a_t$ be non negative integers such that the following properties are satisfied: for any $(x_1, \ldots, x_t) \in C$, one has*

- $\lambda(x_1, \ldots, x_t) \geq 0$,
- *for any $\lambda \in \mathbb{N}$ such that $\lambda < \lambda(x_1, \ldots, x_t)$, $(x_1 - \lambda a_1, x_2 - \lambda a_2, \ldots, x_t - \lambda a_t) \in C'$.*

*Let $p$ be a quasi-polynomial over $C'$ and define the map $F$ as:*

$$F(x_1, \ldots, x_t) = \sum_{0 \,\leq\, \lambda \,\leq\, \lambda(x_1, \ldots, x_t)} p(x_1 - \lambda a_1, x_2 - \lambda a_2, \ldots, x_t - \lambda a_t).$$

*Then $F$ is a quasi-polynomial over $C$.*

*Proof.* The proof of Lemma 5 is the same of that of Lemma 4 except the point we describe now. In the sum above that defines the map $F$, the index $\lambda$ runs over the set of integers of the closed interval $[0, \lambda(x_1, \ldots, x_t)]$ so that:

$$\lambda \leq \lfloor \lambda(x_1, \ldots, x_t) \rfloor.$$

Therefore, in order to prove the claim, one has to prove a slightly modified version of Eq. (9) of Lemma 4, that is: for any $j = 0, \ldots, d-1$,

$$\left\lfloor \frac{\lfloor \lambda(x_1, \ldots, x_t) \rfloor - j}{d} \right\rfloor,$$

is a quasi-polynomial with rational coefficients in the variables $x_1, \ldots, x_t$. This can be done by using an argument very similar to that one adopted in the proof of Eq. (9). □

**Lemma 6** *Let* $\lambda_1, \lambda_2 : \mathbb{N}^t \longrightarrow \mathbb{Q}$ *be 2 maps such that, for any* $(x_1, \ldots, x_t) \in \mathbb{N}^t$*,*

$$\lambda_1(x_1, \ldots, x_t) = b_1 x_1 + \cdots + b_t x_t, \quad \lambda_2(x_1, \ldots, x_t) = c_1 x_1 + \cdots + c_t x_t$$

*where* $b_1, \ldots, b_t$ *and* $c_1, \ldots, c_t$ *are given rational coefficients.*

*Let* $C$ *be a subset of* $\mathbb{N}^t$ *and let* $a_1, \ldots, a_t$ *be non negative integers. Suppose that, for any* $(x_1, \ldots, x_t) \in C$*, one has:*

$$0 \leq \lambda_1(x_1, \ldots, x_t) \leq \lambda_2(x_1, \ldots, x_t),$$

*and, for any* $\lambda \in \mathbb{N}$ *such that* $\lambda_1(x_1, \ldots, x_t) \leq \lambda \leq \lambda_2(x_1, \ldots, x_t)$*, one has:*

$$(x_1 - \lambda a_1, x_2 - \lambda a_2, \ldots, x_t - \lambda a_t) \in C',$$

*where* $C'$ *is a given subset of* $\mathbb{N}^t$*. Let* $p$ *be a quasi-polynomial over* $C'$ *and define the map* $F$ *as:*

$$F(x_1, \ldots, x_t) = \sum_{\lambda_1 \leq \lambda \leq \lambda_2} p(x_1 - \lambda a_1, x_2 - \lambda a_2, \ldots, x_t - \lambda a_t),$$

*where* $\lambda_1 = \lambda_1(x_1, \ldots, x_t)$ *and* $\lambda_2 = \lambda_2(x_1, \ldots, x_t)$*. Then* $F$ *is a quasi-polynomial over* $C$*. The same result holds whenever the index* $\lambda$ *runs in the set of integers of the intervals:*

$$(\lambda_1, \lambda_2), \quad (\lambda_1, \lambda_2], \quad [\lambda_1, \lambda_2).$$

*Proof.* Let us solve the case when $\lambda$ runs in the interval $[\lambda_1, \lambda_2]$. Write

$$F(x_1, \ldots, x_t) = S_1(x_1, \ldots, x_t) - S_2(x_1, \ldots, x_t), \tag{10}$$

where:

$$S_1(x_1, \ldots, x_t) = \sum_{0 \leq \lambda \leq \lambda_2} p(x_1 - \lambda a_1, x_2 - \lambda a_2, \ldots, x_t - \lambda a_t),$$

and

$$S_2(x_1, \ldots, x_t) = \sum_{0 \leq \lambda < \lambda_1} p(x_1 - \lambda a_1, x_2 - \lambda a_2, \ldots, x_t - \lambda a_t).$$

By applying Lemma 5 to $S_1$ and Lemma 4 to $S_2$, we have that $S_1$ and $S_2$ are quasi-polynomial and by Lemma 2, so is $S_1 - S_2$. The claim now follows from Eq. (10). The other three cases are similarly proved. □

**Lemma 7** *Assuming the same hypotheses of Lemma 4, the function*

$$S(x_1, \ldots, x_t) = \sum_{\lambda(x_1, \ldots, x_t) \leq \lambda \leq \lambda(x_1, \ldots, x_t)} p(x_1 - \lambda a_1, x_2 - \lambda a_2, \ldots, x_t - \lambda a_t).$$

*is a quasi-polynomial over* $C$*.*

*Proof.* It is a direct consequence of Lemma 6, assuming $\lambda_1(x_1, \dots, x_t) = \lambda_2(x_1, \dots, x_t) = \lambda(x_1, \dots, x_t)$ □

Now we want to define some suitable regions of $\mathbb{R}^t$. More precisely, our regions will be polyhedral cones determined by a family of hyperplanes passing through the origin. We proceed as follows. Let $\pi$ be a plane of $\mathbb{R}^t$. Let us fix an equation for $\pi$ denoted by $\pi(x) = 0$. We associate with $\pi$ a map

$$f_\pi : \mathbb{R}^t \longrightarrow \{+, -, 0\}$$

defined as: for any $x \in \mathbb{R}^t$,

$$f_\pi(x) = \begin{cases} + & \text{if } \pi(x) > 0, \\ 0 & \text{if } \pi(x) = 0, \\ - & \text{if } \pi(x) < 0. \end{cases}$$

We remark that the map defined above depends upon the plane $\pi$ and its equation in the obvious geometrical way. We can now give the following important two definitions.

**Definition 5** *Let $\Pi = \{\pi_1, \dots, \pi_m\}$ be a family of planes of $\mathbb{R}^t$ that satisfy the following property:*

- $\Pi$ *includes the coordinate planes, that is, the planes defined by the equations $x_\ell = 0,\ \ell = 0, \dots, t$;*
- *every plane of $\Pi$ passes through the origin.*

*Let $\sim$ be the equivalence defined over the set $\mathbb{R}^t$ as: for any $x,\ x' \in \mathbb{R}^t$,*

$$x \sim x' \iff \forall\, i = 1, \dots, m, \quad f_{\pi_i}(x) = f_{\pi_i}(x').$$

*A subset $C$ of $\mathbb{R}^t$ is called a* region (with respect to $\Pi$) *if it is a coset of $\sim$.*

It may be useful to keep in mind that the singleton composed by the origin is a region. Moreover if $t = 2$, the set of all points of $\mathbb{R}^t \setminus \{0\}$ of every line of $\Pi$ is a region also.

**Definition 6** *Let $F : \mathbb{N}^t \longrightarrow \mathbb{N}$ be a map. Then $F$ is said to be a* box spline *in $\mathbb{N}^t$ if there exists a partition $\mathcal{C} = \{C_1, \dots, C_y\}$ of regions of $\mathbb{N}^t$ – defined by a family of planes satisfying Definition 5 – and a family $p_1, \dots, p_y$ of quasi-polynomials, every one of which is associated with exactly a region of $\mathcal{C}$, such that, for any $(x_1, \dots, x_t) \in \mathbb{N}^t$, one has:*

$$F(x_1, \dots, x_t) = p_a(x_1, \dots, x_t),$$

*where $a$ is the index of the region $C_a$ that contains $(x_1, \dots, x_t)$.*

**Lemma 8** *The sum of a finite family of box splines is a box spline.*

*Proof.* It suffices to prove the claim for two box splines. Let $F_1$ and $F_2$ be two box splines and let $\mathcal{C} = \{C_1, \ldots, C_y\}$ and $\mathcal{D} = \{D_1, \ldots, D_z\}$ be the families of regions of $F_1$ and $F_2$ respectively. Moreover, let $\{p_1, \ldots, p_y\}$ and $\{q_1, \ldots, q_z\}$ be the families of quasi polynomials of $F_1$ and $F_2$ respectively.

Consider the box spline defined as follows. Let $\mathcal{E}$ be the partition of regions of $\mathbb{N}^t$ given by the intersection of $\mathcal{C}$ and $\mathcal{D}$ respectively. It is worth noticing that $\mathcal{E}$ is determined by the union of the two families of hyperplanes that define $\mathcal{C}$ and $\mathcal{D}$ respectively. Then we associate the map $r_{lm} = p_l + q_m$ with every region $E_{lm}$ of $\mathcal{E}$. By Lemma 2, $r_{lm}$ is a quasi-polynomial. For any $x \in \mathbb{N}^t$ we have

$$F_1(x) = p_l(x), \quad F_2(x) = q_m(x),$$

where $l$ and $m$ are the indices of the regions $C_l$ and $D_m$ that contain $x$. Hence we have

$$F_1(x) + F_2(x) = r_{lm}(x),$$

while $x$ belongs to the region $E_{lm}$. Since $r_{lm}$ is the quasi polynomial associated with $E_{lm}$, this proves that $F_1 + F_2$ is equal to the box spline defined above. $\square$

**Lemma 9** *Let* $F : \mathbb{N}^{t_1+t_2} \longrightarrow \mathbb{N}$ *be a box spline in* $\mathbb{N}^{t_1+t_2}$*. Then there exists a map* $G : \mathbb{N}^{t_1} \longrightarrow \mathbb{N}$*, which is a box spline in* $\mathbb{N}^{t_1}$*, such that, for every* $(x_1, \ldots, x_{t_1}) \in \mathbb{N}^{t_1}$*, the following equality holds:*
$G(x_1, \ldots, x_{t_1}) = F(x_1, \ldots, x_{t_1}, \underbrace{hx_1, ..., hx_1}_{t_2\text{-times}})$.

*Proof.* Let $\Pi_F = \{\pi_1, \ldots, \pi_m\}$ be the family of planes of $\mathbb{R}^{t_1+t_2}$ associated with $F$.

Recall that $\Pi_F$ satisfies the following properties:

- $\Pi_F$ includes the coordinate planes, that is, the planes defined by the equations $x_\ell = 0,\ \ell = 0, \ldots, t_1 + t_2$;
- every plane of $\Pi_F$ passes through the origin.

Now let $\pi(x_1, \ldots, x_{t_1+t_2}) \equiv \sum_{i=1,\ldots,t_1+t_2} \beta_i x_i = 0$ be a plane in $\Pi_F$. Then the plane:

$$\pi'(x_1, \ldots, x_{t_1}) \equiv \sum_{i=1,\ldots,t_1} \beta_i x_i + \sum_{j=t_1+1,\ldots,t_1+t_2} \beta_j h x_1 = 0$$

is a plane of $\mathbb{R}^{t_1}$ through the origin. We define as $\Pi_G$ the family of all such planes. It is obvious that all coordinates planes belong to $\Pi_G$.

Let now $\bar{x} = (\bar{x}_1, ..., \bar{x}_{t_1})$ be a point in $\mathbb{N}^{t_1}$. Since $F$ is a box-spline, $F$ associates to the point $\bar{x}^* = (\bar{x}_1, ..., \bar{x}_{t_1}, \underbrace{h\bar{x}_1, ..., h\bar{x}_1}_{t_2\text{-times}})$ a unique quasi-polynomial $p(x)$.

We want to show that the region of $\bar{x}$ w.r.t. the planes in $\Pi_G$ determines $p(x)$ univocally. Let $\pi \equiv \sum_{i=1,\ldots,t_1+t_2} \beta_i x_i = 0$ be a plane in $\Pi_F$. Let $\pi'$ be the corresponding plane of $\Pi_G$. Then it is obvious that $\pi(\bar{x}^*) > 0$ (or $\pi(\bar{x}^*) = 0$, or $\pi(\bar{x}^*) < 0$) if and only if $\pi'(\bar{x}) > 0$ (or, respectively, $\pi'(\bar{x}) = 0$, or, respectively, $\pi'(\bar{x}) < 0$). Then the position of $\bar{x}$ w.r.t. $\Pi_G$ determines $p(x)$ univocally.

Now, let $d$ the period of $p(x)$. Observe that a given point $\bar{x} = (\bar{x}_1, ..., \bar{x}_{t_1})$ in $\mathbb{N}^{t_1}$, gives rise to the remainders $(d_1, \ldots, d_{t_1}, \underbrace{d_1^*, ..., d_1^*}_{t_2\text{-times}})$ modulo $d$, where:
$d_1^* = h\bar{x}_1 \operatorname{mod}(d)$.

It follows that, each sequence of remainders $(d_1, \ldots, d_{t_1})$ modulo $d$ specifies the unique polynomial $p_{(d_1,d_2,\cdots,d_{t_1},d_1^*,\ldots,d_1^*)}$ of the quasi-polynomial $p(x)$.

Therefore we let correspond to $p(x)$ the quasi-polynomial $q(x) : \mathbb{N}^{t_1} \longrightarrow \mathbb{N}$, which has period $d$ and to each sequence of of remainders $(d_1, \ldots, d_{t_1})$ modulo $d$ associates the polynomial $q_{(d_1,d_2,\cdots,d_{t_1})} : \mathbb{N}^{t_1} \longrightarrow \mathbb{N}$, with rational coefficients, defined by;

$$q_{(d_1,d_2,\cdots,d_{t_1})}(x_1, ..., x_{t_1}) = p_{(d_1,d_2,\cdots,d_{t_1},d_1^*,\ldots,d_1^*)}(x_1, ..., x_{t_1}, hx_1, ..., hx_1)$$

where $(x_1, ..., x_{t_1})$ is such that $d_i = x_i \operatorname{mod}(d)$, for every $i = 1, ..., t_1$, and $d_1^* = hx_1 \operatorname{mod}(d)$.

The box-spline $G$ is therefore completely specified and this ends the proof. □

The following lemma is a crucial tool in the proof of the main result of this section.

**Lemma 10** *Let $G : \mathbb{N}^t \longrightarrow \mathbb{N}$ be a box spline and let $a_1, \ldots, a_t \in \mathbb{N}$ with $(a_1, ..., a_t) \neq (0, ..., 0)$. Consider the map $\Lambda : \mathbb{N}^t \longrightarrow \mathbb{N}$ that associates with every $(x_1, \ldots, x_t) \in \mathbb{N}^t$, the value*

$$\Lambda(x_1, \ldots, x_t) = \min\left\{ \frac{x_i}{a_i} \mid a_i \neq 0 \right\}.$$

*Let $S : \mathbb{N}^t \longrightarrow \mathbb{N}$ be the map defined as: for every $(x_1, \ldots, x_t) \in \mathbb{N}^t$,*

$$\mathcal{S}(x_1, \ldots, x_t) = \sum_{0 \leq \lambda \leq \Lambda(x_1,\ldots,x_t)} G(x_1 - \lambda a_1, x_2 - \lambda a_2, \ldots, x_t - \lambda a_t). \tag{11}$$

*Then $S$ is a box spline.*

*Proof.* In order to prove the claim, we first associate with the map $S$ a new family of regions that we define now. Let $\Pi$ be the family of planes associated with the box spline $G$. For any $(x_1, x_2, \ldots, x_t) \in \mathbb{N}^t$ consider the line defined by the equation parameterized by $\lambda$:

$$(x_1 - \lambda a_1, x_2 - \lambda a_2, \ldots, x_t - \lambda a_t). \tag{12}$$

Let $\pi$ be a plane of the family $\Pi$ and let $\pi(x) = \sum_{i=1,\ldots,t} \ \beta_i x_i = 0$ be its equation. The value of $\lambda$ that defines the point of meeting of the line (12) with $\pi$ is easily computed. Indeed, $\lambda$ is such that

$$\sum_{i=1,\ldots,t} \beta_i(x_i - \lambda a_i) = 0,$$

so that

$$\sum_{i=1,\ldots,t} \beta_i x_i = \lambda \cdot \sum_{i=1,\ldots,t} \beta_i a_i \tag{13}$$

which gives

$$\lambda = \sum_{i=1,\ldots,t} \frac{\beta_i}{\gamma} x_i, \tag{14}$$

where

$$\gamma = \sum_{i=1,\ldots,t} \beta_i a_i.$$

It is worth to remark that Eq. (14) is not defined whenever

$$\gamma = \sum_{i=1,\ldots,t} \beta_i a_i = 0. \tag{15}$$

Let us first treat Eq. (15). Here, either the line of Eq. (12) belongs to $\pi$ or such a line is parallel to $\pi$. Therefore, for every point $x$ of the line of Eq. (12), the value of $f_\pi(x)$ is constant so that $\pi$ is not relevant in determining a change of region when a point is moving on the line of Eq. (12). Because of this remark, we shall consider only planes of $\Pi$ for which Eq. (15) does not hold. Denote $\Pi'$ this set of planes. For any $\pi \in \Pi'$, with equation $\pi(x) = \sum_{i=1,\ldots,t} \ \beta_i x_i = 0$, consider the homogeneous linear polynomial

$$\lambda_\pi(x_1, \ldots, x_t) = \sum_{i=1,\ldots,t} \frac{\beta_i}{\gamma} x_i,$$

where $\gamma = \sum_{i=1,\ldots,t} \beta_i a_i$. We remark that for any $(x_1, x_2, \ldots, x_t) \in \mathbb{N}^t$, the line parameterized by Eq. (12) meets the plane $\pi$ in the point corresponding to the parameter $\lambda = \lambda_\pi(x_1, \ldots, x_t)$.

Consider an enumeration of the planes of the set $\Pi$ and denote by $<$ the linear order on $\Pi$ defined by such enumeration. Consider the new family $\widehat{\Pi}$ of planes defined by the following sets of equations:

1. $\pi(x) = 0$, $\pi \in \Pi$
2. $\lambda_{\pi\pi'}(x_1, \ldots, x_k) = 0$, with $\pi, \pi' \in \Pi'$, $\pi < \pi'$, and $\lambda_{\pi\pi'}(x_1, \ldots, x_k) = \lambda_\pi(x_1, \ldots, x_k) - \lambda_{\pi'}(x_1, \ldots, x_k)$.

Call $\widehat{\mathcal{C}}$ the family of regions of $\mathbb{N}^t$ defined by $\widehat{\Pi}$.

We now associate with every region of $\widehat{\mathcal{C}}$ a quasi-polynomial. In order to do this, we need to establish some preliminary facts. Let us fix now a region $C$ of $\widehat{\mathcal{C}}$ and let $x = (x_1, \dots, x_t)$ be a point of $\mathbb{N}^t$ that belongs to $C$. Let $i$ be such that

$$\Lambda(x) = \frac{x_i}{a_i}.$$

Observe that, for any other point $x' = (x'_1, \dots, x'_t)$ in $C$, one has

$$\Lambda(x') = \frac{x'_i}{a_i}.$$

Indeed, it is enough to prove that, for any given pair of distinct indices $i, j$, we have:

$$\frac{x_i}{a_i} \leq \frac{x_j}{a_j} \iff \frac{x'_i}{a_i} \leq \frac{x'_j}{a_j}.$$

This is equivalent to say that:

$$\lambda_{\pi\pi'}(x_1, \dots, x_t) \leq 0 \iff \lambda_{\pi\pi'}(x'_1, \dots, x'_t) \leq 0,$$

where $\pi, \pi'$ are the planes $x_i = 0$ and $x_j = 0$ respectively. The previous equivalence is true because $x$ and $x'$ belong to the same region of $\widehat{\mathcal{C}}$.

Another important fact is the following. Let us consider any point $x$ of the region $C$ of $\widehat{\mathcal{C}}$. Consider the subset of planes of $\Pi'$:

$$\{\pi_1, \dots, \pi_m\} = \{\pi \in \Pi' \mid 0 \leq \lambda_\pi(x) \leq \Lambda(x)\}.$$

We can always assume, possibly changing the enumeration of the above planes, that

$$0 \leq \lambda_{\pi_1}(x) \leq \cdots \leq \lambda_{\pi_m}(x) \leq \Lambda(x).$$

**Remark.** Observe that, for any other point $x'$ of $C$, one has

$$\{\pi_1, \dots, \pi_m\} = \{\pi \in \Pi' \mid 0 \leq \lambda_\pi(x') \leq \Lambda(x)\}.$$

and

$$0 \leq \lambda_{\pi_1}(x') \leq \cdots \leq \lambda_{\pi_m}(x') \leq \Lambda(x').$$

The remark above can be proved by using an argument very similar to that used to prove the previous condition. We suppose that the above inequalities are strict, i.e. $0 < \lambda_{\pi_1}(x) < \cdots < \lambda_{\pi_m}(x) < \Lambda(x)$. In this case, as before, one proves that the same inequalities are strict for any other point $x'$ of the region $C$. The case when the inequalities are not all strict can be treated similarly.

From now on, by the sake of clarity, for any $x = (x_1, \dots, x_t) \in \mathbb{N}$ we set $y_\lambda(x) = (x_1 - \lambda a_1, \dots, x_t - \lambda a_t)$.
Consider the following sets:

- $Y_0(x) = \{y_\lambda(x) \mid \lambda \in \mathbb{N} \cap [0, \lambda_{\pi_1}(x))\}$,
- $Y_m(x) = \{y_\lambda(x) \mid \lambda \in \mathbb{N} \cap (\lambda_{\pi_m}, \Lambda(x))\}$,
- $Y_i(x) = \{y_\lambda(x) \mid \lambda \in \mathbb{N} \cap (\lambda_{\pi_i}(x), \lambda_{\pi_{i+1}}(x))\},\ i = 1, \ldots, m-1$.
- $Z_i(x) = \{y_\lambda(x) \mid \lambda \in \mathbb{N} \cap \{\lambda_{\pi_i}(x)\}\},\ i = 1, \ldots, m$.
- $Z_{m+1}(x) = \{y_\lambda(x) \mid \lambda \in \mathbb{N} \cap \{\Lambda(x)\}\}$.

We are now able to associate a quasi-polynomial with the region $C$ of $\widehat{\mathcal{C}}$. For this purpose, take two points $x, x'$ in $C$. By the facts discussed before, one has that the lines of Eq. (12) associated with $x$ and $x'$ respectively, meet the planes of $\Pi'$ in the same order. We recall, that a change of region on the generic line of Eq. (12) happens only when the line meets a plane of $\Pi'$. Therefore, since $\widehat{\mathcal{C}}$ is a refinement of $\mathcal{C}$ and $x$ and $x'$ are in a same region with respect to $\mathcal{C}$, the above conditions imply that, for every $i = 0, \ldots, m$, the two sets of points $Y_i(x)$ and $Y_i(x')$ are subsets of a same common region of $\mathcal{C}$. Hence there exists a quasi-polynomial $p_i$, depending on $i$ and on the region $C$, such that, for any $y \in Y_i(x)$ and for any $y' \in Y_i(x')$, $G(y) = p_i(y), G(y') = p_i(y')$. By the previous remark and by Lemma 6, one has that, for any $i = 0, ..., m$, there exists a quasi-polynomial $q_i$, depending on $i$, and on $C$, such that for any $x \in C$

$$q_i(x) = \sum_{y \in Y_i(x)} G(y).$$

Observe that, since $x$ and $x'$ are in the same region $C$, as before one derives that $Z_i(x)$ and $Z_i(x')$ are in the same region with respect to $\mathcal{C}$. Therefore, as before, by applying Lemma 7 there exists a quasi-polynomial $r_i$, depending on $i$ and on $C$, such that for any $x \in C$

$$r_i(x) = \sum_{y \in Z_i(x)} G(y).$$

On the other hand, by Eq. (11), we have that, for any $(x_1, \ldots, x_t) \in C$, $\mathcal{S}(x_1, \ldots, x_t)$ is equal to:

$$q_0(x) + r_1(x) + q_1(x) + r_2(x) + q_2(x) + \cdots r_m(x) + q_m(x) + r_{m+1}(x). \quad (16)$$

Thus, $\mathcal{S}(x_1, \ldots, x_t)$ on the region $C$ is represented as a sum of quasi-polynomials. This, together with Lemma 2 applied to Eq. (16) imply that the map $S$ is a quasi-polynomial over every region of $\widehat{\mathcal{C}}$. The proof of the claim is thus complete. □

**Theorem 2** *Let*

$$\mathcal{S} : \mathbb{N}^t \longrightarrow \mathbb{N}$$

*be the function which counts, for any vector $(n_1, \ldots, n_t) \in \mathbb{N}^t$, the number of distinct non-negative solutions of a given Diophantine system:*

$$\begin{cases} a_{11}x_1 + a_{12}x_2 + \cdots + a_{1k}x_k = & n_1 \\ a_{21}x_1 + a_{22}x_2 + \cdots + a_{2k}x_k = & n_2 \\ \quad\quad\quad\quad \cdot & \cdot \\ \quad\quad\quad\quad \cdot & \cdot \\ \quad\quad\quad\quad \cdot & \cdot \\ a_{t1}x_1 + a_{t2}x_2 + \cdots + a_{tk}x_k = & n_t. \end{cases} \tag{17}$$

*where the numbers $a_{ij} \in \mathbb{N}$ and, for every $i = 1, \ldots, k$, there exists $j = 1, \ldots, t$ such that $a_{ij} \neq 0$. The function $\mathcal{S}$ is a box spline. Moreover such box spline can be effectively constructed starting from the coefficients of the system.*

*Proof.* For any vector $(n_1, \ldots, n_t) \in \mathbb{N}^t$, let $\mathrm{Sol}(n_1, \ldots, n_t)$ be the set of the non-negative solutions of the Diophantine system (17) and denote by $\mathcal{S} : \mathbb{N}^t \longrightarrow \mathbb{N}$, the map defined as: for any vector $(n_1, \ldots, n_t) \in \mathbb{N}^t$,

$$\mathcal{S}(n_1, \ldots, n_t) = \mathrm{Card}(\mathrm{Sol}(n_1, \ldots, n_t)),$$

that is, it associates with every vector $(n_1, \ldots, n_t)$ the number of non-negative distinct solutions of the system (17). Let us prove that the map $\mathcal{S}$ is a box spline. For this purpose, we proceed by induction on the number of unknowns of the system (17). We start by proving the basis of the induction. In this case, our system has one unknown, say $x$, and it can be written as:

$$\begin{cases} a_1x = & n_1 \\ a_2x = & n_2 \\ \quad \cdot & \\ \quad \cdot & \\ \quad \cdot & \\ a_tx = & n_t \end{cases}$$

The system has solutions (and, in this case, it is unique) if and only if there exists $\lambda \in \mathbb{N}$ such that:

$$\lambda(a_1, \ldots, a_t) = (\lambda a_1, \ldots, \lambda a_t) = (n_1, \ldots, n_t). \tag{18}$$

Let us consider the line $\ell$ (through the origin) defined by the parametric equation (18). The line $\ell$ can be determined as the intersection of suitable planes through the origin. Let us consider the family of regions defined by the set of these planes together with the coordinate planes. One can easily associate with every region a quasipolynomial. For this purpose, we remark that the set of points of the line $\ell$ with integral coordinates, without the origin, is a region. On this region, the counting function of the system takes the value 0 or 1. Therefore this map coincides with the quasi-polynomial given by $p = 0,\ q = 1$ with the periodical rule $d = lcm\{a_1, \ldots, a_t\}$. To any other region, we associate $p$. The basis of the induction is thus proved.

Let us now prove the inductive step. If $(x_1, \ldots, x_k) \in \mathrm{Sol}(n_1, \ldots, n_t)$, the system (17) can be written as:

$$\begin{cases} a_{12}x_2 + \cdots + a_{1k}x_k = & n_1 - a_{11}x_1 \\ a_{22}x_2 + \cdots + a_{2k}x_k = & n_2 - a_{21}x_1 \\ \quad\quad\quad\cdot & \quad\quad\cdot \\ \quad\quad\quad\cdot & \quad\quad\cdot \\ \quad\quad\quad\cdot & \quad\quad\cdot \\ a_{t2}x_2 + \cdots + a_{tk}x_k = & n_t - a_{t1}x_1. \end{cases} \tag{19}$$

This implies that:

$$n_1 - a_{11}x_1 \geq 0, \;\; n_2 - a_{21}x_1 \geq 0, \;\; n_t - a_{t1}x_1 \geq 0,$$

so that, since $x_1$ must be an integer $\geq 0$, one has:

$$0 \leq x_1 \leq \frac{n_1}{a_{11}}, \;\; 0 \leq x_1 \leq \frac{n_2}{a_{21}}, \quad \ldots, \;\; 0 \leq x_1 \leq \frac{n_t}{a_{t1}},$$

and thus:

$$0 \leq x_1 \leq \Lambda(n_1, \ldots, n_t),$$

where the map $\Lambda : \mathbb{N}^t \longrightarrow \mathbb{N}$ is defined as:

$$\Lambda(x_1, \ldots, x_t) = \min\left\{ \frac{x_i}{a_{i1}} \;\mid\; a_{i1} \neq 0 \right\}. \tag{20}$$

We remark that, since the vector $(a_{11}, a_{21}, \ldots, a_{t1}) \neq (0, 0, \ldots, 0)$, the map $\Lambda$ is well defined. Set $K = \lfloor \Lambda(x_1, \ldots, x_t) \rfloor$. We can write $\mathrm{Sol}(n_1, \ldots, n_t)$ as:

$$\mathrm{Sol}(n_1, \ldots, n_t) = (0 \times \mathrm{Sol}_0) \cup (1 \times \mathrm{Sol}_1) \cup \ldots \cup (K \times \mathrm{Sol}_K), \tag{21}$$

where, for every $i = 0, \ldots, K$, $\mathrm{Sol}_i$ denotes the set of non-negative solutions of the Diophantine system:

$$\begin{cases} a_{12}x_2 + \cdots + a_{1k}x_k = & n_1 - a_{11}i \\ a_{22}x_2 + \cdots + a_{2k}x_k = & n_2 - a_{21}i \\ \quad\quad\quad\cdot & \quad\quad\cdot \\ \quad\quad\quad\cdot & \quad\quad\cdot \\ \quad\quad\quad\cdot & \quad\quad\cdot \\ a_{t2}x_2 + \cdots + a_{tk}x_k = & n_t - a_{t1}i. \end{cases} \tag{22}$$

By Eq. (21), for any $(n_1, \ldots, n_t) \in \mathbb{N}^t$, we have:

$$\mathcal{S}(n_1, \ldots, n_t) = \sum_{i=0,\ldots,K} \mathrm{Card}(\mathrm{Sol}_i). \tag{23}$$

By applying the inductive hypothesis to the system (22), we have that there exists a box spline $G : \mathbb{N}^t \longrightarrow \mathbb{N}$ such that, for any $(n_1, \ldots, n_t) \in \mathbb{N}^t$, if $0 \leq i \leq K$,

$$\mathrm{Card}(\mathrm{Sol}_i) = G(n_1 - a_{11}i, n_2 - a_{21}i, \ldots, n_t - a_{t1}i), \tag{24}$$

so that, by Eq. (23) and Eq. (24), one has:

$$\mathcal{S}(n_1,\ldots,n_t) = \sum_{0\le\lambda\le\Lambda(x_1,\ldots,x_t)} G(n_1-\lambda a_{11}, n_2-\lambda a_{21},\ldots,n_t-\lambda a_{t1}). \quad (25)$$

By Eq. (25), the fact that $\mathcal{S}$ is a box spline follows from Lemma 10. Finally we remark that the proof gives an effective procedure to construct the claimed box spline that describes the map $\mathcal{S}$. □

**Corollary 1** *Let*

$$\mathcal{S} : \mathbb{N}^t \longrightarrow \mathbb{N}$$

*be the function which counts, for any vector* $(n_1,\ldots,n_t) \in \mathbb{N}^t$*, the number of distinct non-negative solutions of a given Diophantine system:*

$$\begin{cases} a_{10} + a_{11}x_1 + a_{12}x_2 + \cdots + a_{1k}x_k = & n_1 \\ a_{20} + a_{21}x_1 + a_{22}x_2 + \cdots + a_{2k}x_k = & n_2 \\ \cdot & \cdot \\ \cdot & \cdot \\ \cdot & \cdot \\ a_{t0} + a_{t1}x_1 + a_{t2}x_2 + \cdots + a_{tk}x_k = & n_t, \end{cases} \quad (26)$$

*where the numbers* $a_{ij} \in \mathbb{N}$ *and, for every* $i = 1,\ldots,k$*, there exists* $j = 1,\ldots,t$ *such that* $a_{ij} \neq 0$*. Set* $a_0 = (a_{10}, a_{20},\ldots,a_{t0})$*. Then, on the set of all vectors* $x \in \mathbb{N}^t$ *with* $x \ge a_0$*, the map* $\mathcal{S}$ *is a box spline. Moreover such box spline can be effectively constructed starting from the coefficients of the system.*

*Proof.* First consider the system

$$\begin{cases} a_{11}x_1 + a_{12}x_2 + \cdots + a_{1k}x_k = & n_1 \\ a_{21}x_1 + a_{22}x_2 + \cdots + a_{2k}x_k = & n_2 \\ \cdot & \cdot \\ \cdot & \cdot \\ \cdot & \cdot \\ a_{t1}x_1 + a_{t2}x_2 + \cdots + a_{tk}x_k = & n_t. \end{cases} \quad (27)$$

According to Theorem 2, there exists a box spline $F$ that counts, for every $(n_1,\ldots,n_t) \in \mathbb{N}^t$, the number of the solutions of the diophantine system (27). Let $\mathcal{C} = \{C_1,\ldots,C_s\}$ be the partition of $\mathbb{N}^t$ in regions and let $\{p_1,\ldots,p_s\}$ be the family of quasi-polynomials that define $F$.

Let $a_0 = (a_{10},\ldots,a_{t0})$ be the vector whose components are the entries of the first column of the matrix of the system (26). For every $\eta \in \mathbb{N}^t$ with $\eta \ge a_0$, the components of the vector $\eta - a_0$ are non negative integers so that:

$$\mathcal{S}(\eta) = p_j(\eta - a_0),$$

where $j$ is the index of the region of the family $\mathcal{C}$ that contains the vector $\eta - a_0$. This concludes the proof. □

**Example 1** For the sake of clarity, we find useful to show the proof of Theorem 2 on the following very simple example. Consider the Diophantine system:

$$\begin{cases} x_1 + 2x_2 = & n_1 \\ 2x_1 + 3x_2 = & n_2, \end{cases} \tag{28}$$

where $n_1, n_2 \in \mathbb{N}$ and let $F : \mathbb{N}^2 \longrightarrow \mathbb{N}$ be the counting function of the system (28). By following the proof of Theorem 2, together with that of Lemma 10, we construct the partition of $\mathbb{N}^2$ in regions and the family of quasi-polynomials that describe the function $F$. In the sequel, the following notation is adopted: $(x_1, x_2)$ and $(n_1, n_2)$ are respectively the vector of the unknowns and the vector of the non homogeneous terms of the system, while $x, y$ are free variables over the set $\mathbb{N}$. Observe that $x_1, x_2$ gives a solution of (28) if and only if $(n_1 - x_1, n_2 - 2x_1) = (2t, 3t),\ t \geq 0$.

Therefore consider the Diophantine system:

$$\begin{cases} 2x_2 = & n_1 \\ 3x_2 = & n_2 \end{cases} \tag{29}$$

where $n_1, n_2 \in \mathbb{N}$. Let $G : \mathbb{N}^2 \longrightarrow \mathbb{N}$ be the counting function of the system (29). Let $\Pi = \{\pi_1, \pi_2, \pi_3\}$ be the set of the lines defined by the equations:

$$\pi_1(x,y) \equiv x = 0,\ \pi_2(x,y) \equiv y = 0,\ \pi_3(x,y) \equiv 3x - 2y = 0.$$

Let $\mathcal{R}$ be the partition of $\mathbb{N}^2$ determined by $\Pi$. Then $\mathcal{R}$ is formed by the following 6 regions:

$R_0 = \{(0,0)\},\ R_1 = \{(x,0)\ :\ x > 0\},\ R_2 = \{(x,y)\ :\ 3x > 2y,\ x, y > 0\},$

$R_3 = \{(x,y)\ :\ 3x = 2y,\ \ x, y > 0\},\ R_4 = \{(x,y)\ :\ 3x < 2y,\ x, y > 0\},$

$R_5 = \{(0,y)\ :\ y > 0\}.$

Let $\mathcal{P}$ be the family of polynomials given by:

$p_0(x,y) = p_3(x,y) \equiv 1,\ p_1(x,y) = p_2(x,y) = p_4(x,y) = p_5(x,y) \equiv 0.$

One can check that the box spline determined by $\mathcal{R}$ and $\mathcal{P}$ is the function $G$.

Let $(x,y)$ be a given point of $\mathbb{N}^2$ and let $\ell(\lambda)$ be the line represented by the equation parameterized by $\lambda$:

$$(x - \lambda, y - \lambda).$$

For every $i = 1, 2, 3$, let $\lambda_{\pi_i}(x,y)$ be the value of $\lambda$ that defines the point of meeting of the line $\ell(\lambda)$ with $\pi_i$. Then one has:

$$\lambda_{\pi_1}(x,y) = x,\ \lambda_{\pi_2}(x,y) = y/2,\ \lambda_{\pi_3}(x,y) = 2y - 3x.$$

Consider the new family $\widehat{\Pi}$ of lines defined by the following sets of equations:

1. $\pi_i(x,y)=0,\ i=1,2,3,$ (that is, the lines in $\Pi$),

2. For every pair $i,j$ of indices with $1\le i<j\le 3$, let

$$\lambda_{\pi_{ij}}(x,y)\equiv\lambda_{\pi_i}(x,y)-\lambda_{\pi_j}(x,y)=0.$$

It easily checked that there exists exactly one line of the previous form (2) which is represented by the equation $y-2x=0$. Thus one has:

$$\widehat{\Pi}=\{x=0,\ y=0,\ 2x-3y=0,\ 2x-y=0\}.$$

If we denote by $\widehat{\mathcal{R}}$ the family of regions of $\mathbb{N}^2$ defined by $\widehat{\Pi}$, we have:

$\widehat{R}_0=\{(0,0)\},\ \widehat{R}_1=\{(x,0)\ :\ x>0\},\ \widehat{R}_2=\{(x,y)\ :\ 3x>2y,\ y>0\},$

$\widehat{R}_3=\{(x,y)\ :\ 3x=2y,\ \ y>0\},\ \widehat{R}_4=\{(x,y)\ :\ 3x<2y,\ 2x>y,\ x,y>0\},$

$\widehat{R}_5=\{(x,y)\ :\ 2x=y,\ y>0\},\ \widehat{R}_6=\{(x,y)\ :\ 2x<y,\ x,y>0\},$

$\widehat{R}_7=\{(0,y)\ :\ y>0\}.$

Now we associate with every region $\widehat{R}_i$ of $\widehat{R}$ a quasi-polynomial $\hat{q}_i$. Actually, set:

$$\hat{q}_0=\hat{q}_3=\hat{q}_4=\hat{q}_5\equiv 1 \text{ and } \hat{q}_1=\hat{q}_2=\hat{q}_6=\hat{q}_7\equiv 0.$$

One can check that the box spline determined by $\widehat{\mathcal{R}}$ together with the list of polynomials above is the function $F$.

# 4 The growth function of a semi-linear set

The main result of this section is that the growth function of a semi-linear set is a box-spline. Let us start by introducing a definition.

**Definition 7** *Let $X$ be a subset of $\mathbb{Z}^t$ and let $t_1,t_2\in\mathbb{N}$ such that $t=t_1+t_2$. We associate with $X$ a function*

$$\mathcal{G}^+{}_{X,\ t_1,t_2}:\mathbb{N}^t\longrightarrow\mathbb{N}$$

*that returns, for every $(n_1,\ldots,n_{t_1},m_1\ldots,m_{t_2})\in\mathbb{N}^t$, the number of elements*

$(x_1, \ldots, x_t) \in X$ *such that:*

$$\begin{cases} x_1 = n_1 \\ \vdots \\ x_{t_1} = n_{t_1} \\ 0 \leq x_{t_1+1} \leq m_1 \\ \vdots \\ 0 \leq x_t \leq m_{t_2}. \end{cases} \tag{30}$$

*The function* $\mathcal{G}^+{}_{X,t_1,t_2}$ *is called the* (restricted) generalized growth function *of* $X$*.*

**Theorem 3** *Let* $X$ *be a semi-linear set of* $\mathbb{N}^t$ *and let* $t_1, t_2 \in \mathbb{N}$ *such that* $t = t_1 + t_2$*. Then* $\mathcal{G}^+{}_{X,t_1,t_2}$ *is a box spline.*

*Proof.* Let $t_1, t_2 \in \mathbb{N}$ be given as above and let $X$ be a semi-linear set of $\mathbb{N}^t$. To avoid a heavy notation, in this proof, we suppress the dependency of $\mathcal{G}^+{}_X$ on $t_1$ and $t_2$ and we simply write $\mathcal{G}_X$. By Theorem 1, $X$ is semi-simple so that

$$X = \bigcup_{i=1,\ldots,\ell} X_i$$

is a finite and disjoint union of simple sets of $\mathbb{N}^t$. As a straightforward consequence, one can easily check that, for every $(n_1, ..., n_t) \in \mathbb{N}^t$,

$$\mathcal{G}_X(n_1, ..., n_t) = \sum_{1}^{\ell} \mathcal{G}_{X_i}(n_1, ..., n_t).$$

In order to obtain the claim, by the equality above and by Lemma 8 it is enough to prove that, for every simple set $Y$ of $\mathbb{N}^t$, $\mathcal{G}_Y$ is a box spline.
For this purpose, let

$$Y = b_0 + b_1^{\oplus} + \cdots + b_k^{\oplus},$$

be a representation of $Y$ as a simple set where, for every $j = 0, ..., k$,

$$b_j = (b_{j1}, \cdots, b_{jt}) \in \mathbb{N}^t.$$

Therefore we can write Eq. (30) for the set $Y$ as a system of $t$ inequalities in $k$ unkowns $y_1, ..., y_k$:

$$\begin{cases} b_{01} + b_{11}y_1 + b_{21}y_2 + \cdots + b_{k1}y_k & = n_1 \\ b_{02} + b_{12}y_1 + b_{22}y_2 + \cdots + b_{k2}y_k & = n_2 \\ . & . \\ . & . \\ . & . \\ b_{0t_1} + b_{1t_1}y_1 + b_{2t_1}y_2 + \cdots + b_{kt_1}y_k & = n_{t_1} \\ \\ b_{0t_1+1} + b_{1t_1+1}y_1 + b_{2t_1+1}y_2 + \cdots + b_{kt_1+1}y_k & \leq m_1 \\ b_{0t_1+2} + b_{1t_1+2}y_1 + b_{2t_1+2}y_2 + \cdots + b_{kt_1+2}y_k & \leq m_2 \\ . & . \\ . & . \\ . & . \\ b_{0t} + b_{1t}y_1 + b_{2t}y_2 + \cdots + b_{kt}y_k & \leq m_{t_2}. \end{cases} \tag{31}$$

Since $Y$ is a simple set of $\mathbb{N}^t$, there exists a bijection between the set of non-negative solutions of the previous system and the set of elements of $Y$ that satisfy Eq. (30). Consider now the Diophantine system of equations obtained from that of Eq. (31) where $z_1, z_2, ..., z_{t_2}$ form a set of $t$ unknowns disjoint from the set of unknowns $y_1, ..., y_k$:

$$\begin{cases} b_{01} + b_{11}y_1 + b_{21}y_2 + \cdots + b_{k1}y_k & = n_1 \\ b_{02} + b_{12}y_1 + b_{22}y_2 + \cdots + b_{k2}y_k & = n_2 \\ . & . \\ . & . \\ . & . \\ b_{0t_1} + b_{1t_1}y_1 + b_{2t_1}y_2 + \cdots + b_{kt_1}y_k & = n_{t_1} \\ \\ b_{0t_1+1} + b_{1t_1+1}y_1 + b_{2t_1+1}y_2 + \cdots + b_{kt_1+1}y_k + z_1 & = m_1 \\ b_{0t_1+2} + b_{1t_1+2}y_1 + b_{2t_1+2}y_2 + \cdots + b_{kt_1+2}y_k + z_2 & = m_2 \\ . & . \\ . & . \\ . & . \\ b_{0t} + b_{1t}y_1 + b_{2t}y_2 + \cdots + b_{kt}y_k + z_{t_2} & = m_{t_2}. \end{cases} \tag{32}$$

One can easily check that the number of non-negative solutions of the Diophantine system of inequalities of Eq. (31) is equal to the number of non-negative solutions of the Diophantine system of Eq. (32). By Corollary 1 applied to the latter system, we have that its the counting function is a box spline. By the previous facts, the function $\mathcal{G}_Y$ is a box spline as well and this concludes the proof. □

Now we consider a first extension of Theorem 3 to semi-linear sets over $\mathbb{Z}^t$. For this purpose the following lemma is useful (see [2]).

**Lemma 11** *If $X$ is linear in $\mathbb{Z}^t$ then $X \cap \mathbb{N}^t$ is semi-linear in $\mathbb{N}^t$.*

**Corollary 2** *If $X$ is semi-linear in $\mathbb{Z}^t$ then $X \cap \mathbb{N}^t$ is semi-linear in $\mathbb{N}^t$.*

*Proof.* We can write $X$ as a finite union of linear sets $X_1, ..., X_\ell$ of $\mathbb{Z}^t$. Thus we have

$$X \cap \mathbb{N}^t = \bigcup_{i=1,\dots,\ell} X_i',$$

where, for every $i = 1, ..., \ell$, $X_i' = X_i \cap \mathbb{N}^t$. The claim now follows from the fact that, by Lemma 11, for every $i = 1, ..., \ell$, $X_i$ is semi-linear in $\mathbb{N}^t$. □

**Corollary 3** *Let $X$ be a semi-linear set of $\mathbb{Z}^t$ and let $t_1, t_2 \in \mathbb{N}$ such that $t = t_1 + t_2$. Set $X' = X \cap \mathbb{N}^t$. Then $\mathcal{G}_{X',\ t_1,t_2}$ is a box spline.*

*Proof.* Since, by hypotheses, $X$ is semi-linear in $\mathbb{Z}^t$, by Corollary 2, $X'$ is semi-linear in $\mathbb{N}^t$. The claim follows by applying Theorem 3 to $X'$. □

**Definition 8** *Let $X$ be a subset of $\mathbb{Z}^t$ and let $t_1, t_2 \in \mathbb{N}$ such that $t = t_1 + t_2$. We associate with $X$ a function*

$$\mathcal{G}_{X,\ t_1,t_2} : \mathbb{N}^t \longrightarrow \mathbb{N}$$

*that returns, for every $(n_1, \dots, n_{t_1}, m_1 \dots, m_{t_2}) \in \mathbb{N}^t$, the number of elements $(x_1, \dots, x_t) \in X$ such that:*

$$\begin{cases} |x_1| & = & n_1 \\ \cdot & & \cdot \\ \cdot & & \cdot \\ \cdot & & \cdot \\ |x_{t_1}| & = & n_{t_1} \\ |x_{t_1+1}| & \leq & m_1 \\ \cdot & & \cdot \\ \cdot & & \cdot \\ \cdot & & \cdot \\ |x_t| & \leq & m_{t_2}. \end{cases} \tag{33}$$

*The function $\mathcal{G}_{X,\ t_1,t_2}$ is called the* generalized growth function *of $X$.*

**Remark 5** If $t_1 = 0$, we obtain the function defined in Eq. (1) of Definition 2 that we have called the growth function of $X$.

**Theorem 4** *Let $X$ be a semi-linear set of $\mathbb{Z}^t$ and let $t_1, t_2 \in \mathbb{N}$ such that $t = t_1 + t_2$. Then $\mathcal{G}_{X,\ t_1,t_2}$ is a box spline.*

In order to prove the theorem, we need some preliminaries. First we refresh some notions given in Section 2. Let $\Pi$ be the family of planes $\pi_i$ of equation $x_i = 0$, for every $i = 1, ..., t$ and let $\sim$ be the equivalence relation introduced in Definition 5. We recall that a region is a coset of $\mathbb{Z}^t$ with respect to $\sim$. Let $\mathcal{C}$ be the family of regions associated with $\Pi$. It is easily checked that every region $C$ is a semilinear set of $\mathbb{Z}^t$. On the other hand, we recall that the family of semi-linear sets of $\mathbb{Z}^t$ is closed with respect to the Boolean set operations. These facts allows one to obtain the following useful result.

**Lemma 12** *Let $X$ be a semi-linear set of $\mathbb{Z}^t$ and let $C \in \mathcal{C}$. Then the set $X \cap C$ is still semi-linear in $\mathbb{Z}^t$.*

**Lemma 13** *Let $X$ be a semi-linear set of $\mathbb{Z}^t$ and let $t_1, t_2 \in \mathbb{N}$ such that $t = t_1 + t_2$. Let $C \in \mathcal{C}$ and set $Y = X \cap C$. There exists a bijection*

$$f : \mathbb{Z}^t \longrightarrow \mathbb{Z}^t,$$

*that satisfies the following conditions:*

1. *$f(Y) \subseteq \mathbb{N}^t$,*
2. *the map $f$ preserves semi-linearity in $\mathbb{Z}^t$,*
3. *For every $\eta \in \mathbb{N}^t$, $\mathcal{G}_{Y,\ t_1,t_2}(\eta) = \mathcal{G}^+{}_{f(Y),\ t_1,t_2}(\eta)$.*

*Therefore, the map $\mathcal{G}_{Y,\ t_1,t_2}$ is a box spline.*

*Proof.* For the sake of simplicity, we study the case $t = 2$, the general case being treated similarly. Set $\mathbb{N}_+ = \mathbb{N} \setminus \{0\}$. The family $\mathcal{C}$ is given by:

- $C_1 = \{(0, 0)\}$,
- $C_2 = \mathbb{N}_+^2$,
- $C_3 = \{(-x, y)\} \mid x, y \in \mathbb{N}_+\}$,
- $C_4 = \{(-x, -y)\} \mid x, y \in \mathbb{N}_+\}$,
- $C_5 = \{(x, -y)\} \mid x, y \in \mathbb{N}_+\}$,
- $C_6 = \{(0, y)\} \mid y \in \mathbb{N}_+\}$,
- $C_7 = \{(0, -y)\} \mid y \in \mathbb{N}_+\}$,
- $C_8 = \{(x, 0)\} \mid x \in \mathbb{N}_+\}$,
- $C_9 = \{(-x, 0)\} \mid x \in \mathbb{N}_+\}$.

Set $Y = X \cap C$, where $C \in \mathcal{C}$ and $X$ is a semi-linear set of $\mathbb{Z}^2$. If $C = C_1$ then the result is trivial. If $C \in \{C_2, C_6, C_8\}$, then the result follows by taking $f$ as the identity on $\mathbb{Z}^2$ and by applying Corollary 3 to $\mathcal{G}^+{}_{Y,\ t_1,t_2}(\eta)$. Assume $C = C_3$ and define the function $f : \mathbb{Z}^2 \longrightarrow \mathbb{Z}^2$ as: for every $(z_1, z_2) \in \mathbb{Z}^2$

$$f(z_1, z_2) = (-z_1, z_2).$$

One immediately has $f(Y) \subseteq \mathbb{N}^2$ which gives Condition (1). Moreover, the function $f$ is an isometry which gives Condition (3). Let us prove Condition (2). If $X$ is a linear set of $\mathbb{Z}^t$, $t \geq 2$, and $B$ is a representation of $X$, then one can easily check that $f(B)$ is a representation of $f(X)$. Then Condition (2) follows from the fact that a semilinear set is a finite union of linear sets. Finally the fact that the function $\mathcal{G}_{Y,\ t_1,t_2}$ is a box spline follows from Condition (3) by applying Corollary 3 to $\mathcal{G}^+{}_{f(Y),\ t_1,t_2}$. We remark that the other two cases $C_4, C_5, C_7, C_9$ can be treated similarly as $C_3$. □

We are now in position to prove Theorem 4.

**Proof of Theorem 4** Let $X$ be a semi-linear set of $\mathbb{Z}^t$ and let $t_1, t_2 \in \mathbb{N}$ such that $t = t_1 + t_2$. The set $X$ can be written as a finite and disjoint union

$$X = \bigcup_{C \in \mathcal{C}} (X \cap C).$$

Therefore, for any $\eta \in \mathbb{N}^t$, one has:

$$\mathcal{G}_{X,\ t_1,t_2}(\eta) = \sum_{C \in \mathcal{C}} \mathcal{G}_{(X \cap C,\ t_1,t_2)}(\eta). \tag{34}$$

If $C$ is any region of $\mathbb{Z}^t$, the fact that $\mathcal{G}_{X \cap C,\ t_1,t_2}$ is a box spline follows from Lemma 13. Finally the claim follows by applying Lemma 8 to Eq. (34). □

# 5 A combinatorial proof of a theorem of Dahmen and Micchelli.

Let $A$ be a matrix in $\mathbb{Z}^{t \times n}$, with $n \geq t$. Assume that the following condition holds:

$$\forall\ X \in \mathbb{Z}^n,\ X \geq 0^n,\ AX = 0^t \implies X = 0^n. \tag{35}$$

The following property holds.

**Lemma 14** *Let $A$ be a matrix in $\mathbb{Z}^{t \times n}$, with $n \geq t$, satisfying (35). If $b$ is a vector in $\mathbb{Z}^t$, then the number of non-negative integer solutions of the system $AX = b$ is always finite.*

*Proof.* By contradiction, assume that the number of non-negative integer solutions of the system $AX = b$ is infinite. Since $\mathbb{N}^n$ is well quasi-ordered, there exist two solutions $X_1$ and $X_2$ of the system $AX = b$ such that $X_1 > X_2$. Thus the vector $X_1 - X_2$ is a non null non-negative solution of the system $AX = 0^t$. This contradicts the hypothesis that $A$ satisfies (35). The claim is thus proved. □

Therefore, given a matrix $A$ satisfying (35), we can define a function

$$\mathcal{C}_A : \mathbb{Z}^t \longrightarrow \mathbb{N},$$

which associates, with every vector $b \in \mathbb{Z}^t$, the number of non-negative integer solutions of the system $AX = b$. The following problem has been addressed and solved in [5].

**Theorem 5** *(Dahmen and Micchelli,1988) Let $A$ be a matrix in $\mathbb{Z}^{t\times n}$, with $n \geq t$, that satisfies (35). Then the function $\mathcal{C}_A$ is a box spline in $\mathbb{Z}^t$.*

In the following, we give a new proof of the latter theorem. Our proof is based on the results of the previous sections and on the properties of semi-simple sets in $\mathbb{Z}^t$.

Let $S_{A,b}$ be the set of all non-negative real solutions of the system $AX = b$, with $b \in \mathbb{Z}^t$. Now, we want to show that Condition (35) on the matrix $A$ implies that $S_{A,b}$ is a bounded set in $\mathbb{R}^n_+$. First, as done in [5], we observe that (35) is equivalent to say that the convex hull $H_A$ of the columns $a_i$'s of the matrix $A$, *that is,*

$$H_A = \left\{ \big(\sum_{i=1}^{n} \alpha_i a_{ji}\big)_{j=1,\ldots,t} \;\middle|\; \alpha_i \in \mathbb{R},\, \alpha_i \geq 0,\, \sum_{i=1}^{n} \alpha_i = 1 \right\},$$

does not contain the origin $0^t$.

Let $X$ be a closed convex set of $\mathbb{R}^t$ and let $x$ be a point of the boundary of $X$. An hyperplane $\pi$ such that $x \in \pi$ and $X$ is contained in one of the two closed halfspaces determined by $\pi$, is called *support hyperplane* of $X$ at point $x$. The following well-known result of Convex geometry holds.

**Theorem 6** *([11], Theorem 4.1) Given a closed convex subset $X$ of $\mathbb{R}^t$ and a point $x$ of the boundary of $X$, there exists a support hyperplane of $X$ at $x$, not necessarily unique.*

Denote by $||x||$ the Euclidean norm of a point $x \in \mathbb{R}^t$. Let $A$ be a matrix in $\mathbb{Z}^{t\times n}$, with $n \geq t$, that satisfies (35). Hence we have $0^t \notin H_A$. Since $H_A$ is a convex and closed set of $\mathbb{R}^t$, by applying Theorem 6 to $H_A$, there exists a hyperplane $\pi$ that separates $H_A$ from the origin $0^t$. If $\delta$ is the minimal distance of $0^t$ from $\pi$, then $\delta > 0$ so that we have:

**Corollary 4** *Let $A$ be a matrix in $\mathbb{Z}^{t\times n}$, with $n \geq t$, that satisfies (35). Then there exists a real number $\delta > 0$ such that, for every $x \in H_A$, $||x|| > \delta$.*

**Proposition 1** *Let $A$ be a matrix in $\mathbb{Z}^{t\times n}$, with $n \geq t$, that satisfies (35). If $b$ is a vector in $\mathbb{Z}^t$, then the set $S_{A,b}$ is bounded in $\mathbb{R}^n_+$.*

*Proof.* We argue by contradiction. Assume that $S_{A,b}$ is not bounded. Then there exists a partition of $\{1,\ldots,n\}$ in two sets $I$ and $J$ such that the following property holds: for every $i \in I$ (resp. $j \in J$), the set of values that appear in every vector of $S_{A,b}$ at position $i$ (resp. $j$) is unbounded (resp. bounded). Let $i_0$ be in $I$. We can define an infinite sequence of solutions $(X^r)_{r\geq 1}$ which is unbounded and such that for every $r$, $X^r_{i_0} < X^{r+1}_{i_0}$. Consider now any other coordinate $i_1 \neq i_0$ in $I$. If the set of all $\{X^r_{i_1}\}_{r\geq 1}$, is bounded then we do nothing. Otherwise we extract from the sequence $(X^r)_{r\geq 1}$ a subsequence $(X'^r)_{r\geq 1}$, which is unbounded on the coordinate $i_1$ also and such that, for every $r \geq 1$, $X'^r_{i_1} < X'^{r+1}_{i_1}$. By applying this argument finitely many times, we can obtain a subset $I'$ of $I$ and a sequence $\sigma = (Y^r)_{r\geq 1}$ of solutions such that:

- for every $i \in I'$, the set $\{\sigma^r_i\}_{r\geq 1}$ is unbounded;
- for every $r$ and for every $i \in I'$, $Y^r_i < Y^{r+1}_i$;
- for every $i \in J' = \{1,\ldots,n\} - I'$, the set $\{\sigma^r_i\}_{r\geq 1}$ is bounded;

Therefore we can extract from the sequence $\sigma$ a subsequence $(\widehat{Y}^r)_{r\geq 1}$ such that, for every coordinate $i \in J'$, the sequence $(\widehat{Y}^r_i)_{r\geq 1}$ is convergent. Denote by $||v||$ the Euclidean norm of a vector $v$. By the latter condition, for every $\epsilon > 0$, we can find two positive integers $j < j'$, such that:

$$\forall\ i \in J',\ \ ||\widehat{Y}^{j'}_i - \widehat{Y}^j_i|| < \epsilon. \tag{36}$$

Let us consider the vector $Z$ such that:

$$\forall\ i \in J',\ Z_i = 0, \quad \forall\ i \in J',\ Z_i = \widehat{Y}^{j'}_i - \widehat{Y}^j_i. \tag{37}$$

By Eq. (36), one can easily check that $||AZ|| < c\epsilon$ for some fixed constant $c$. Observe now that since $\sigma$ is unbounded on each coordinate $i \in I'$ we can assume that $Z_i > 1$, for every $i \in I'$. Let $D = \sum_{i'\in I'} Z'_i$. It is easily seen that, by replacing in Eq. (37), each $Z_i$, $i \in J'$, with $Z_i/D$, we get a new vector $Z$ such that the inequality $||AZ|| < Dc\epsilon$ still holds. Since $AZ \in H_A$ and $\epsilon$ is arbitrary, this contradicts Corollary 4. This ends the proof.

Now we want to determine a suitable hypercube which includes the set $S_{A,b}$. To this aim we consider the following *linear programming* that we call problem (Max $A, b$):

- maximize $X_i$, where $i$ is a given index, subject to:
- $AX = b$, and
- $X \geq 0^n$.

We make use of the following theorem (*cf* [12], Theorem 2.2)

**Theorem 7** *Let $B$ be a matrix in $\mathbb{Z}^{t\times n}$, with $n \geq t$ and let $f, d$ be integer vectors. Consider the linear programming problem in standard form given by:*

- *minimize $d \cdot X = \sum_{i=1}^{n} d_i X_i$, subject to:*
- *$BX = f$, and*
- *$X \geq 0^n$.*

*If an optimal solution exists then there exists also an optimal solution $X$ such that for every $i$, with $i = 1, \ldots, n$, $|X_i| \leq h\beta$, where $h$ is a non-negative integer constant, depending only on the matrix $B$ and the vector $d$, and $\beta = \max\{|f_j|, |d_i|\}$.*

We can now apply the previous theorem to solve Problem (Max $A, b$).

**Corollary 5** *Let $A$ be a matrix in $\mathbb{Z}^{t\times n}$, with $n \geq t$, that satisfies (35). If $b$ is a vector in $\mathbb{Z}^t$, then there exists a non-negative integer constant $h_A$, depending only on the matrix $A$, such that the set $S_{A,b}$ is contained in the hypercube $X_i \leq h_A\beta$, for $i = 1, \ldots, n$, where $\beta = \max\{|b_j|\}$, with $j = 1, \ldots, t$.*

*Proof.* By Proposition 1, the set $S_{A,b}$ is bounded in $\mathbb{R}^n_+$ so that Problem (Max $A, b$) admits an optimal solution. Then the claim follows by applying Theorem 7 with $B = A$, $f = b$ and choosing any vector $d$ such that $d \in \{0, -1\}^n$ with exactly one non-null component. □

Now we give a new proof of Theorem 5. For this purpose, we need some preliminary lemmas. The system $AX = b$ is written in the form:

$$\left\{\begin{array}{rl} a_{11}x_1 + a_{12}x_2 + \cdots + a_{1n}x_n = & b_1 \\ a_{21}x_1 + a_{22}x_2 + \cdots + a_{2n}x_n = & b_2 \\ \cdot & \cdot \\ \cdot & \cdot \\ \cdot & \cdot \\ a_{t1}x_1 + a_{t2}x_2 + \cdots + a_{tn}x_n = & b_t. \end{array}\right. \tag{38}$$

Let $\Lambda : \mathbb{Z}^t \longrightarrow \mathbb{N}$ be the map defined as:

$$\forall\ b = (b_1, \ldots, b_t) \in \mathbb{Z}^t,\ \Lambda(b) = h_A \max_{1\leq i\leq t} |b_i|,$$

where $h_A$ is the positive integer defined in Corollary 5. Consider the family $\Pi$ of (hyper-)planes defined by the following equations:

- for every $i = 1, \ldots, t$, $x_i = 0$,
- for every $i, j$ with $1 \leq i < j \leq t$, $x_i - x_j = 0$ and $x_i + x_j = 0$,

and let $\mathcal{R}$ be the family of regions of $\mathbb{Z}^t$ defined by $\Pi$. The first two lemmas easily derive from the definition of $\mathcal{R}$.

**Lemma 15** *For any region $R$ of $\mathcal{R}$, there exist $\alpha_1, \ldots, \alpha_t \in \{\pm 1\}$, and $j$ with $1 \leq j \leq t$, such that, for every $b = (b_1, \ldots, b_t) \in R$:*

- $\forall\ i = 1, \ldots, t,\ |b_i| = (-1)^{\alpha_i} b_i,$
- $\Lambda(b) = (-1)^{\alpha_j} h_A b_j.$

**Remark 6** Lemma 15 shows the role of the family $\Pi$. Indeed, if $b = (b_1, \ldots, b_t)$ is a vector of $\mathbb{Z}^t$, for every components $b_i, b_j$ of $b$, one can determine whether $|b_i| = b_i$ or $|b_i| = -b_i$ and whether $|b_i| \leq |b_j|$ or $|b_i| > |b_j|$.

**Lemma 16** *Let $R$ be a region of $\mathcal{R}$. There exist $\alpha_1, \ldots, \alpha_t \in \{\pm 1\}$ such that, for any $b = (b_1, \ldots, b_t) \in R$, the system (38) is equivalent to the following:*

$$\begin{cases} |a_{11}x_1 + a_{12}x_2 + \cdots + a_{1n}x_n| = & |b_1| \\ |a_{21}x_1 + a_{22}x_2 + \cdots + a_{2n}x_n| = & |b_2| \\ \cdot & \cdot \\ \cdot & \cdot \\ \cdot & \cdot \\ |a_{t1}x_1 + a_{t2}x_2 + \cdots + a_{tn}x_n| = & |b_t|. \\ (-1)^{\alpha_1}(a_{11}x_1 + a_{12}x_2 + \cdots + a_{1n}x_n) \geq 0 & \\ (-1)^{\alpha_2}(a_{21}x_1 + a_{22}x_2 + \cdots + a_{2n}x_n) \geq 0 & \\ \cdot & \cdot \\ \cdot & \cdot \\ \cdot & \cdot \\ (-1)^{\alpha_t}(a_{t1}x_1 + a_{t2}x_2 + \cdots + a_{tn}x_n) \geq 0 & \\ x_1 \geq 0 & \\ x_2 \geq 0 & \\ \cdot & \\ \cdot & \\ \cdot & \\ x_n \geq 0 & \end{cases} \tag{39}$$

**Lemma 17** *Let $R$ be a region of $\mathcal{R}$. There exists a semilinear set $Z$ of $\mathbb{Z}^{t+n}$ such that, for every $b = (b_1, \ldots, b_t) \in R$:*

$$\mathcal{C}_A(b) = G(|b_1|, \ldots, |b_t|, \Lambda(b), \ldots, \Lambda(b)),$$

*where $G \equiv \mathcal{G}_{Z,\ t_1,t_2}$ is the generalized growth function of $Z$ with $t_1 = t$ and $t_2 = n$.*

*Proof.* Let $R$ be a region of $\mathcal{R}$. By Lemma 16, there exist $\alpha_1, \ldots, \alpha_t \in \{\pm 1\}$ such that, for any $b = (b_1, \ldots, b_t) \in R$, the system (38) is equivalent to a system of the form (39). Let $Z$ be the subset of all elements of $\mathbb{Z}^{t+n}$ defined as:

$$Z = \{(z_1, \ldots, z_{t+n})\ :\ z_i \geq 0,\ i = 1, \ldots, t+n\},\ \text{where}$$

- $\forall\ i = 1, \ldots, t,\ z_i = (-1)^{\alpha_i}(a_{i1}x_1 + a_{i2}x_2 + \cdots + a_{in}x_n)$,
- $\forall\ i = t+1, \ldots, t+n,\ z_i = x_{i-t}$.

It is easily seen that $Z$ is semilinear in $\mathbb{Z}^{t+n}$. Let $G \equiv \mathcal{G}_{Z,\ t_1,t_2}$ be the generalized growth function of $Z$ with $t_1 = t$ and $t_2 = n$. From the definition of $Z$ and by applying Corollary 5 to (38), for every $b \in R$, one has $\mathcal{C}_A(b) = G(|b_1|, \ldots, |b_t|, \Lambda(b), \ldots, \Lambda(b))$. □

**Lemma 18** *Let $R$ be a region of $\mathcal{R}$. The function $\mathcal{C}_A$ is a box spline on $R$.*

*Proof.* Let $R$ be a region of $\mathcal{R}$. By Lemma 16, there exists an index $j_0$, with $1 \leq j_0 \leq t$ such that for every $b = (b_1, \ldots, b_t) \in R$, $\Lambda(b) = h_A|b_{j_0}|$.

By Lemma 17, there exists a semilinear set $Z$ of $\mathbb{Z}^{t+n}$ such that, for every $b = (b_1, \ldots, b_t) \in R$ :

$$\mathcal{C}_A(b) = G(|b_1|, \ldots, |b_t|, \Lambda(b), \ldots, \Lambda(b)) = G(|b_1|, \ldots, |b_t|, h_A|b_{j_0}|, \ldots, h_A|b_{j_0}|), \tag{40}$$

where $G$ is the generalized growth function of $Z$ with $t_1 = t$ and $t_2 = n$. By applying Theorem 4 to $G$, one has that $G$ is a box spline in $\mathbb{N}^{t+n}$.

By Lemma 9, there exists a box spline $G'$ in $\mathbb{N}^t$ such that, for every $b = (b_1, \ldots, b_t) \in R$ :

$$G'(|b_1|, \ldots, |b_t|) = G(|b_1|, \ldots, |b_t|, h_A|b_{j_0}|, \ldots, h_A|b_{j_0}|). \tag{41}$$

We want to show that there exists a $F : \mathbb{Z}^t \longrightarrow \mathbb{Z}$, which is a box spline in $\mathbb{Z}^t$, such that for every $b = (b_1, \ldots, b_t) \in R$ the following equality holds:

$$F(b_1, \ldots, b_t) = G'(|b_1|, \ldots, |b_t|). \tag{42}$$

First of all, recall that by lemma 15 there exist $\alpha_1, \ldots, \alpha_t \in \{\pm 1\}$, such that for every $b = (b_1, \ldots, b_t) \in R$:

$$\forall\ i = 1, \ldots, t,\ |b_i| = (-1)^{\alpha_i} b_i. \tag{43}$$

Let $\Pi_{G'} = \{\pi_1, \ldots, \pi_m\}$ be the family of planes of $\mathbb{R}^t$ associated with $G'$.

Recall that $\Pi_{G'}$ satisfies the following properties:

- $\Pi_{G'}$ includes the coordinate planes, that is, the planes defined by the equations $x_\ell = 0,\ \ell = 0, \ldots, t$;
- every plane of $\Pi_{G'}$ passes through the origin.

Now let $\pi(x_1, \ldots, x_t) \equiv \sum_{i=1,\ldots,t}\ \beta_i x_i = 0$ be a plane in $\Pi_{G'}$. Then the plane:

$$\pi'(x_1, \ldots, x_t) \equiv \textstyle\sum_{i=1,\ldots,t}\ (-1)^{\alpha_i}\beta_i x_i = 0$$

is a plane of $\mathbb{R}^t$ through the origin. We define as $\Pi_F$ the family of all such planes. It is obvious that all coordinates planes belong to $\Pi_F$.

Let now $\bar{x} = (\bar{x}_1, ..., \bar{x}_t)$ be a point in $R$. Since $G'$ is a box-spline, $G'$ associates to the point $\bar{x}^* = ((-1)^{\alpha_1}\bar{x}_1, ..., (-1)^{\alpha_t}\bar{x}_t)$ a unique quasi-polynomial $p(x)$.

We want to show that the region of $\bar{x}$ w.r.t. the planes in $\Pi_F$ determines $p(x)$ univocally. Let $\pi \equiv \sum_{i=1,\ldots,t} \beta_i x_i = 0$ be a plane in $\Pi_{G'}$. Let $\pi'$ be the corresponding plane of $\Pi_F$. Then it is obvious that $\pi(\bar{x}^*) > 0$ (or $\pi(\bar{x}^*) = 0$, or $\pi(\bar{x}^*) < 0$) if and only if $\pi'(\bar{x}) > 0$ (or, respectively, $\pi'(\bar{x}) = 0$, or, respectively, $\pi'(\bar{x}) < 0$). Then the position of $\bar{x}$ w.r.t. $\Pi_F$ determines $p(x)$ univocally.

Now, let $d > 0$ the period of $p(x)$. If a given point $\bar{x} = (\bar{x}_1, ..., \bar{x}_t)$ in $\mathbb{N}^t$, gives rise to the remainders $(d_1, \ldots, d_t)$ modulo $d$, then $\bar{x}^*$ determine the same remainders $(d_1, \ldots, d_t)$ modulo $d$.

Therefore we let correspond to $p(x)$ the quasi-polynomial $q(x) : \mathbb{Z}^t \longrightarrow \mathbb{Z}$, which has period $d$ and to each sequence of of remainders $(d_1, \ldots, d_t)$ modulo $d$ associates the polynomial $q_{(d_1,d_2,\cdots,d_t)} : \mathbb{Z}^t \longrightarrow \mathbb{Z}$, with rational coefficients, defined by:

$$q_{(d_1,d_2,\cdots,d_t)}(x_1, ..., x_t) = p_{(d_1,d_2,\cdots,d_t)}((-1)^{\alpha_1}x_1, ..., (-1)^{\alpha_t}x_t)$$

The box-spline $F$ is therefore completely specified and this ends the proof. □

As a corollary of the previous lemma, we obtain.

**Theorem 8** *The function $\mathcal{C}_A$ of the Diophantine system (38) is a box spline in $\mathbb{Z}^{t+n}$.*

*Proof.* By Lemma 18, the function $\mathcal{C}_A$ is a box spline on every region of $\mathcal{R}$. From the latter fact, one derives that $\mathcal{C}_A$ is a box spline on $\mathbb{Z}^{t+n}$. □

**Acknowledgements**. Special thanks to Corrado De Concini for very useful comments and discussions concerning the results presented in this paper.